\documentclass[a4paper,fleqn,usenatbib]{mnras}

\usepackage{newtxtext,newtxmath}
\usepackage[T1]{fontenc}
\usepackage{ae,aecompl}

\usepackage{graphicx}	
\usepackage{amsmath}	

\newcommand{\OIII}{\mbox{[O\,{\sc iii}]}}

\newcommand{\HI}{\mbox{H\,{\sc i}}}
\newcommand{\HII}{\mbox{H\,{\sc ii}}}

\newcommand{\kms}{km~s$^{-1}$}

\newcommand{\ergs}{erg~s$^{-1}$}
\newcommand{\cmmt}{cm$^{-2}$}
\newcommand{\cmt}{cm$^{-3}$}

\newcommand{\NH}{\mbox{$\Sigma_{\rm H}$}}

\newcommand{\rdust}{\mbox{$r_{\rm dust}$}}
\newcommand{\Lbol}{\mbox{$L_{\rm bol}$}}

\newcommand{\Lfive}{\mbox{$L_{5 \rm GHz}$}}
\newcommand{\Lfourtyfive}{\mbox{$L_{45 \rm GHz}$}}
\newcommand{\Lhundred}{\mbox{$L_{100 \rm GHz}$}}
\newcommand{\LOIII}{\mbox{$L_{\rm [O~\textsc{iii}]}$}}

\title[RPC. V. Free-free absorption and emission]{Radiation pressure confinement -- V. The predicted free-free absorption
and emission in active galactic nuclei}

\author[A.~Baskin and A.~Laor]{
	Alexei Baskin$^{1}$\thanks{E-mail: alexeiba@soreq.gov.il (AB); laor@physics.technion.ac.il (AL)}
	and Ari Laor$^{2}$\footnotemark[1]
	\\
	$^{1}$Plasma Physics Department, Soreq Nuclear Research Center, Yavne~8180000, Israel\\
	$^{2}$Physics Department, Technion -- Israel Institute of Technology, Haifa~3200000, Israel
}

\date{Accepted 2021 September 7. Received 2021 August 21; in original form 2021 April 13}

\pubyear{2021}

\begin{document}
	\label{firstpage}
	\pagerange{\pageref{firstpage}--\pageref{lastpage}}
	\maketitle

	\begin{abstract}

The effect of radiation pressure compression (RPC) on ionized gas in Active Galactic Nuclei (AGN)
likely sets the photoionized gas density structure. 
The photoionized gas free-free 
absorption and emission are therefore uniquely set by the incident ionizing flux. 
We use the photoionization code Cloudy RPC model results to derive
the expected relations between the free-free emission and 
absorption properties and the distance from 
the AGN centre, for a given AGN luminosity. The free-free absorption frequency of RPC gas 
is predicted to increases from $\sim$100~MHz on the kpc scale, to $\sim$100~GHz on the sub-pc scale,
consistent with observations of spatially resolved free-free absorption. 
The free-free emission at 5~GHz is predicted to yield a radio loudness
of $R\sim 0.03$, below the typical observed values of $R\sim 0.1-1$ 
in radio-quiet AGN. However, the flat free-free radio continuum may become dominant above 100~GHz. 
The suggested detection of optically thin 
free-free emission in NGC~1068, on the sub pc torus
scale, is excluded as the brightness temperature is too high for optically thin free-free 
emission. However, excess
emission observed with ALMA above 150~GHz in NGC~1068, is consistent with the 
predicted free-free emission from gas
just outside the broad line region, a region which overlaps the hot dust 
disc resolved with GRAVITY. Extended $\sim$100~pc scale free-free emission is also likely present in 
NGC~1068. Future sub mm observation of radio quiet AGN
with ALMA may allow to image the free-free emission of warm photoionized gas in 
AGN down to the 30 mas scale, including highly absorbed AGN.    
	\end{abstract}

	\begin{keywords}
		galaxies: active -- quasars: general -- radio continuum: galaxies
	\end{keywords}

	\section{Introduction}\label{sec:intro}

Free-free absorption of radio emission is observed in some AGN \citep[e.g.][]{OS21}, 
and is sometimes spatially resolved, 
allowing to map the distribution of the absorbing gas density and column density 
\citep{V94, W94, LLV95, U99,U99a, G99, Ped99, W00, Jones01, Middel04}. Free-free radio emission
is expected from photoionized gas in AGN \citep{Ulvestad81, KL89}, and is possibly
detected in some AGN \citep{Antonucci88, Barvainis96, Mundell00, G04, C19}. 
The density and ionization structure of photoionized gas in AGN are likely set 
by the incident ionizing radiation energy and momentum fluxes (see below),
and one can therefore derive the expected free-free absorption and emission of the gas just from its
distance from the AGN centre. The purpose of this study
is to provide detailed predictions of these properties, which can provide new constraints 
on photoionized gas in AGN, using the high angular resolution of the mm arrays, and the
great penetration power of the mm radiation. Below, we briefly review the Radiation Pressure Compression (RPC) 
mechanism, which sets both the gas density structure and the gas ionization structure.
	
Radiation deposits both energy and momentum when it interacts with gas. 
The energy deposition and the resulting gas emission received much attention, 
and is calculated by highly detailed photoionization modelings (e.g.\ \citealt{F11}).
The radiation momentum deposition, specifically the radiation force per unit mass, 
is commonly considered significant if it becomes comparable
to gravity. This is relevant from the stellar structure scale \citep{Eddington16}, out to the 
galaxy structure scale \citep[e.g.][]{Thompson05, Murray10}. Radiative acceleration 
can dominate the gas dynamics, and whether the system remains gravitationally bound. 
However, another critical parameter is the ratio of the radiation pressure to the
other internal forces in the gas, in particular the gas pressure. If the gas
is in free fall, say in a Keplerian orbit, then gravity is zero in the gas frame, 
and the only remaining net forces (in the absence of significant magnetic fields)
are the radiation flux gradient and the gas pressure gradient within the gas. 

The energy deposition of the incident ionizing radiation and the gas cooling set the 
gas equilibrium temperature.
The gas density in photoionization modelings is often considered a free parameter. 
However, if the gas pressure gradient is lower than the opposing incident radiation pressure gradient, 
then radiation pressure inevitably compresses the gas. This compression builds up the gas pressure, until 
a local equilibrium is achieved. The built up gas pressure at the depth where the incident radiation 
is fully absorbed, matches then the incident radiation pressure.
Thus, the gas density is not a free parameter, 
as the gas pressure, and also its temperature, are set by the incident radiation pressure. 
This RPC
effect implies that the gas temperature structure and the density structure are both 
set by the incident radiation. The RPC effect was described in a few studies 
\citep{PV95,B98,D02,D11b}, in the context of gas in the
Narrow Line Region (NLR) in AGN and gas in \HII\ regions. The RPC effect has profound implications on the emission and absorption 
line properties of gas in AGN. It likely produces the observed density distance relation of $n\propto r^{-2}$
in the NLR  \citep*{paperI}, the relative strength and the radial dependence of the 
lines from the
Broad Line Region (BLR; \citealt*{paperII}). It also explains the universal absorption measure distribution 
observed in AGN over
a wide range of ionization states  
\citep*{paperIII}, and why the fast outflows in broad absorption line quasars are
not overionized \citep*{paperIV}. The RPC effect also produces the observed universal differential emission measure of the X-ray narrow lines in AGN \citep{paperV}.

The photoionized gas cools by line and continuum emission, in particular 
free-free emission. The purpose of this study is to use the 
RPC modelings to find the expected radio free-free absorption and emission signature 
of RPC gas in AGN. As we show below, the free-free absorption optical depth provides the distance
of the absorbing gas from the centre. The radio free-free emission may be the dominant emission 
mechanism in radio quiet AGN in the mm range. 
The unique advantage of the mm emission is that it is effectively unabsorbable. 
It can penetrate any column of neutral gas, and also the expected ionized column of photoionized gas.
The sensitivity of current mm arrays 
may allow to image the free-free emission sources, 
and thus image photoionized gas, also in highly absorbed AGN, 
on scales of $\sim$100~mas and below, which is not achievable in other bands.

 The paper is organized as follows. In Section~\ref{sec:model} we describe the free-free emission from RPC photoionized gas,
and provide analytic approximations. In Section~\ref{sec:results} we provide the results for a variety of possible configurations. 
The results are compared with observations in Section~\ref{sec:comare_with_obs} and discussed in Section~\ref{sec:discussion}. In Section~\ref{sec:conclusions} we provide 
the main conclusions.

	\section{The Model}\label{sec:model}

\subsection{Simplified estimates}
	
	We show below that RPC implies a nearly constant ratio of free-free  luminosity density
to bolometric luminosity, $L_\nu/\Lbol$, in AGN. This ratio depends only on whether 
the line emitting gas is dusty (NLR), or dustless (BLR), and on the gas covering factor $\Omega$ 
(the fraction of $4\pi$ covered). The relation holds for optically thin emission, 
a condition which can be verified from the observed spectral slope (see below). 

The free-free emissivity is
	\begin{equation}
	 \epsilon_\nu = \frac{2^5 \pi e^6}{3 m_e c^3}\left(\frac{2 \pi}{3 k m_e}\right)^{1/2} T^{-1/2} Z_i^2 n_e n_i {\rm e}^{-h\nu/k T} \bar{g}_{\rm ff} , \label{eq:em}
	\end{equation}
	where $n_e$ and $n_i$ are the electron and ion number density, for ions of charge $Z_i$, $T$ is the gas temperature, and $\bar{g}_{\rm ff}$ is the velocity averaged Gaunt factor 
(e.g.\ \citealt[eq.~5.14a]{RL04}). Most of the ionizing continuum is absorbed near the H ionization front where $T\sim10^{4}$~K. In the radio $h\nu \ll k T$, and thus \ $e^{-h\nu/k T}\approx 1$. The velocity averaged Gaunt factor is $\bar{g}_{\rm ff}\approx 4$ \citep{D11}.  In the analytic estimates below, we assume for simplicity H-only gas (i.e.\ $Z_i=1$). The approximations above yield an emissivity of
	\begin{equation}
	 \epsilon_\nu \approx 2.7 \times10^{-39} n_e n_i  \mbox{~\ergs~Hz$^{-1}$~cm$^{-3}$}.\label{eq:em1}
	\end{equation}
The RPC effect predicts that the ionized gas is compressed to a density of
	\begin{equation}
	 n\approx 1.8 \Lbol r^{-2}\mbox{~\cmt},\label{eq:n}
	\end{equation}
	 near the H ionization front  (\citealt{paperI}, eq.~6), with a compression length-scale of (\citealt{paperII}, eq.~14)
	\begin{equation}
	 l = \frac{2 k T c}{\bar{\sigma}} \left(\frac{L_{\rm bol}}{4\pi r^2}\right)^{-1} = 1\bar{\sigma}^{-1} L_{\rm bol}^{-1} r^2\text{ cm}, \label{eq:l_scale}
	\end{equation}
	where $\bar{\sigma}\approx 10^{-21}$~\cmmt\ for dusty gas, and $\bar{\sigma}\approx10^{-22}$~\cmmt\ for dustless ionized gas at the highest density near the ionization front. Note that \Lbol\ and $r$ are in cgs units. Equation~\ref{eq:l_scale} implies that $l$ is typically small compared to $r$ with
	\begin{equation}
		\frac{l}{r} = 3\times10^{-7} \sigma_{-21}^{-1} L_{46}^{-1} r_{\rm pc}, \label{eq:l_ref_to_r}
	\end{equation}
	where $\bar{\sigma}\approx 10^{-21} \sigma_{-21}$, $L_{\rm bol}=10^{46}L_{46}$ and $r=1r_{\rm pc}$~pc.
	Thus, the specific free-free luminosity from a photoionized uniform density slab, of thickness $l$, distance $r$ from the centre, with a solid angle $4\pi\Omega$, is approximately	
\begin{equation}
	 L_\nu=4\pi r^2 \Omega l \epsilon_\nu \approx 1 \times10^{-37}\Omega \bar{\sigma}^{-1} \Lbol\mbox{~\ergs Hz$^{-1}$}, 
\label{eq:L_ratio}
	\end{equation}
	which implies
	\begin{equation}
	L_\nu/\Lbol\approx
	\begin{cases}
        10^{-16}\, \Omega \mbox{~Hz$^{-1}$} & \text{for dusty gas},\\
        10^{-15}\, \Omega \mbox{~Hz$^{-1}$}  & \text{for dustless gas}.
        \end{cases} \label{eq:L_nu_toL}
	\end{equation}
	The fraction of bolometric luminosity converted to free-free emission, $L_\nu/\Lbol$, 
depends only on $\Omega$, the covering factor of the reprocessing
gas.  The proportionality of $L_\nu \propto \Lbol$ is a general property of photoionized gas, but the exact quantitative relation depends on the gas density, which sets the gas temperature and ionization state. The new result of RPC is that the proportionality coefficient is uniform in a given AGN, and among all AGN, as the ionization structure in the illuminated gas is universal.

	The threshold value, $\nu_{\rm thick}$, below which the gas becomes optically thick to free-free absorption can be evaluated as follows. The free-free absorption coefficient is
(e.g.\ \citealt[eq.~5.18a]{RL04})
	\begin{equation}
	 \alpha_\nu = \frac{4 e^6}{3m_e hc}\left(\frac{2\pi}{3km_e}\right)^{1/2} T^{-1/2} Z_i^2 n_e n_i \nu^{-3} \left(1-{\rm e}^{-h\nu/kT}\right)\bar{g}_{\rm ff}\ . \label{eq:abs_co}
	\end{equation}
	In the radio, $h\nu\ll k T$ and thus 
	\begin{equation}
	 \alpha_\nu\approx  0.08 T^{-3/2} n_e n_i \nu^{-2}\mbox{~cm$^{-1}$}\ ,
\label{eq:abs_co_approx}
	\end{equation}
	for $Z_i=1$ and $\bar{g}_{\rm ff}\approx 4$. The slab is optically thick when 
$\tau_\nu=\alpha_\nu l > 1$, where $l$ is given by eq.~\ref{eq:l_scale}, which 
for $T=10^4$~K occurs below the frequency
	\begin{equation}
	 \nu_{\rm thick} = \sqrt{8\times 10^{-8} \bar{\sigma}^{-1} \Lbol r^{-2}}\ .
	 \end{equation}
It is convenient to define the quantity
	 \begin{equation}
          \rdust\equiv 0.2L_{46}^{0.5}~{\rm pc}, \label{eq:rdust}
	\end{equation}
    which is roughly the sublimation radius of large graphite grain
(see \citealt{BL18}). Since, $\Lbol =2.63\times 10^{10}r_{\rm dust}^2$ in cgs, we get
	 \begin{equation}
	 \nu_{\rm thick} =
	 \begin{cases}
	  1.5\times10^{12} \left(r/r_{\rm dust}\right)^{-1}\text{ Hz} & \text{for dusty gas,}\\
	  4.6\times10^{12} \left(r/r_{\rm dust}\right)^{-1} \text{ Hz}  & \text{for dustless gas.}
	 \end{cases}
	 \label{eq:v_thin_cases}
	\end{equation}
The free-free emission becomes self-absorbed at $\nu<\nu_{\rm thick}$, and if the gas
is isothermal, the emission approaches a blackbody ($L_{\nu}\propto \nu^2$), in contrast with the flat optically thin 
spectrum ($L_{\nu}\propto \nu^{-0.1}$) at $\nu>\nu_{\rm thick}$.

One should note that the radio emission cannot propagate freely in plasma below the plasma frequency,
	\begin{equation}
	 \nu_{\rm p}=\sqrt{\frac{n e^2}{\pi m_e}} \approx 1.6\times 10^4  L_{\rm bol}^{1/2} r^{-1} \text{ Hz},
	\end{equation}
	where $n$ is taken from eq.~\ref{eq:n}. Equivalently,
	\begin{equation}
	 \nu_{\rm p} \approx 2.6\times 10^9 \left(r/r_{\rm dust}\right)^{-1}  \text{ Hz},
	\label{eq:plasma}
        \end{equation}
which is lower than the above estimates of $\nu_{\rm thick}$ by a factor of $\sim 10^3$. The gas
is highly optically thick at $\nu_{\rm p}$, and its emission approaches a blackbody ($L_{\nu}\propto \nu^2$) 
already at $\nu_{\rm thick}$. Thus, $L_{\nu}$ drops by a factor of $10^6$ from
$\nu_{\rm thick}$ to $\nu_{\rm p}$, and the free-free contribution is  
negligible already above the frequency where the plasma cutoff sets in.

	The above estimate of $L_\nu/\Lbol$ (eq.~\ref{eq:L_nu_toL}) 
applies for optically thin free-free emission, i.e at $\nu>\nu_{\rm thick}$.
Below we derive the free-free emission based on a numerical photoionization 
solution which includes RPC. We use the equation of $\bar{g}_{\rm ff}$ from \citet[section 10.2]{D11}. For convenience we set $\epsilon_\nu=0$ for $\nu<10\nu_{\rm p}$.

	\subsection{The numerical solution} \label{sec:numeric_sol}
	
Below we evaluate the structure of a photoionized RPC gas slab, i.e.\ the solution for $n_e$, $n_i$ and $T$, as a function of depth into the slab. We follow the procedure detailed in \citet{paperII}. The structure is calculated using the photoionization code Cloudy 13.03 \citep{Cloudy_2013}. An open geometry is assumed. We use the Cloudy command `constant pressure' to set the total pressure constant throughout the calculation. We also utilize the command `no radiation pressure' to exclude the radiation pressure due to trapped line emission, as otherwise the Cloudy solution scheme may become unstable \citep{paperII}. We assume gas metallicity of $Z=2Z_{\sun}$, and use the scaling law of metals with $Z$ from \citet{groves_etal04}. The intermediate ionizing spectral slope of $\alpha_{\rm ion}=-1.6$ is adopted from \citet{paperII}, with\ $f_\nu \propto \nu^{\alpha_{\rm ion}}$ between 1~Ryd and 1~keV (912--12~\AA). The calculation is stopped when the H ionized fraction (H$^+$/H) drops to 1~per cent. The solution of an RPC slab is independent of the initial density $n_0$ at the illuminated face of the slab, if $P_{\rm rad}\gg P_{\rm gas}$ at the illuminated face \citep{paperI,paperII}. For the BLR calculations, specifically for regions
at $0.1<r/\rdust\leq1$, we set $n_0=10^4$~\cmt. For regions further inwards 
at $r/\rdust\leq0.1$, we set $n_0=10^6$~\cmt. For regions outside the BLR  
we use $n_0=100$~\cmt, except for dustless calculations at $1<r/\rdust\leq10$ which are set with $n_0=10^4$~\cmt. For region at $r/\rdust > 10^4$ we use $n_0=1$~\cmt. These values
for $n_0$ ensure that $P_{\rm rad}\gg P_{\rm gas}$ holds at the illuminated face, and that the slab total width is significantly smaller compared to $r$ (by a factor of $\ge 10$).

The structure of an RPC slab is set by the incident flux, i.e. by the value of $L_{\rm bol}/r^2$, rather than by $L_{\rm bol}$ and $r$ independently. We therefore construct a grid of models for $L_{46}=1$, varying only $r$. Finally, we assume that the gas is either dustless or dusty. For the dusty-gas models, we assume the Galactic ISM grain composition and metal depletion, and a 
linear scaling of the dust-to-gas ratio with $Z$.
	
The free-free emission spectrum is calculated by using the evaluated structure of the photoionized slab. A Cloudy calculation results in a solution of the slab which is divided into consecutive `zones'. Each zone is defined by its width $\Delta r_{\rm z}$ and distance from the continuum source $r_{\rm z}$. Cloudy solves for the physical parameters $n_e$, $n_i$ and $T$ in each zone. We use these parameters to calculate $\epsilon_\nu$ and $\alpha_\nu$ (eqs.~\ref{eq:em} and \ref{eq:abs_co}, respectively) of each zone. We sum over the contribution of the most abundant elements: H, He, C, N, O and Ne, i.e.\ those which satisfy $X/{\rm H}>10^{-4}$ for $Z=2Z_{\sun}$ (for $Z=Z_{\sun}$, we adopt the solar composition set in Cloudy). We evaluate $\bar{g}_{\rm ff}$ in eqs.~\ref{eq:em} and \ref{eq:abs_co} by adopting the two following relations from \citet{D11}  
\begin{equation}
	\bar{g}_{\rm ff}(\nu) =
		\frac{\sqrt{3}}{\pi}\left[\ln\frac{(2kT)^{3/2}}{\pi Z_i e^2 m_e^{1/2}\nu} -\frac{5\gamma}{2}\right],
		\label{eq:g_ff_hot}
\end{equation}
for $h\nu\leq kT/10^3$, where $\gamma=0.577$ is Euler's constant; and\\
\begin{equation}
	\bar{g}_{\rm ff}(\nu) =		
			\ln\left\{\exp\left[5.960 -   \frac{\sqrt{3}}{\pi}\ln\left(Z_i\frac{\nu}{10^9}\left(\frac{T}{10^4}\right)^{-3/2} \right) \right]+{\rm e}\right\},
	\label{eq:g_ff_cold}
\end{equation} 
otherwise. The difference between the two relations near the transition $\nu$, i.e.\ at $h\nu\approx kT/10^3$, is smaller than 5 per cent. Near the H ionization front, where $T\simeq10^4$~K, both relations imply $\bar{g}_{\rm ff}(\nu)\sim \nu^{-0.10}$ for $\nu\approx 1-10$~GHz. The power-law slope steepens slightly for larger values of $\nu$; and at $\nu\approx100$~GHz, the slope equals  $-0.16$ and $-0.14$ for eqs.~\ref{eq:g_ff_hot} and \ref{eq:g_ff_cold}, respectively.

The specific luminosity produced by a given zone of width $\Delta r_{\rm z}$ inside the slab 
is
\begin{equation}
	L_{\nu,\rm z}^{\rm em} = 4\pi\Omega r_{\rm z}^2 \Delta r_{\rm z} \epsilon_\nu.
\end{equation} 
We account for the effect of free-free absorption by the emitting zone itself by taking the luminosity that is emitted by the zone toward the two neighbouring zones as
\begin{equation}
	L_{\nu,\rm z} =L_{\nu,\rm z}^{\rm em} \exp(-\alpha_\nu\Delta r_{\rm z}/2).
\end{equation}
The specific luminosity of a particular zone that escapes the slab from a given side (either the illuminated or back side) is evaluated as
\begin{equation}
	L_{\nu,\rm z}^{\rm esc} = L_{\nu,\rm z}\exp\left(-\sum_i \alpha_{\nu,i}\Delta r_{{\rm z},i} \right), \label{eq:Lnu_z_esc}
\end{equation}
where the summation is over all zones that are located between the particular zone and the front or back surface. The total emitted spectrum from a given side, which we present below, equals 
\begin{equation}
	L_{\nu} = \sum_{\rm z} L_{\nu,\rm z}^{\rm esc}, \label{eq:Lnu_calc}
\end{equation}
where the summation is over all zones.
	
For the sake of comparison, we also calculate the free-free emission from the commonly used uniform gas density photoionization models. The Cloudy calculations are executed in a similar setting to the RPC calculation, as described above. The main differences are the following. First, the gas density $n$, rather than the total $P$, is kept constant at the assumed value. Second, the density of ionizing photons at the illuminated side of the slab, $n_\gamma$, is set by $n$ and the ionization parameter 
$U\equiv n_\gamma/n$, rather than by $r/\rdust$. Finally, the stopping criterion is either reaching a H ionization fraction of 1 per cent, as for the RPC calculations, or a total H column of $\NH = 5\times10^{24}$~\cmmt\ (RPC solutions always reach the criterion of ionization fraction prior to reaching this \NH). Only the highest $\log U=1$ dustless calculation stops due to the \NH\ criterion (reaching a H ionization fraction of $\sim$1.5 per cent at the back side). All other dustless calculations terminate at the 1 per cent H ionization fraction; and all dusty calculations stop because they reach the default Cloudy lower limit on the temperature of $T=4000$~K. The procedure for evaluating the outgoing free-free emission 
is the same as described above for the RPC models.

\section{Results} \label{sec:results}

\subsection{The gas density and temperature structure} \label{sec:structure}

The common approach when calculating free-free absorption and emission is to assume
a uniform density and temperature structure \citep[e.g.][]{Ulvestad81, KL89}. 
However, as discussed above, the observations suggest that the gas 
density distribution in the NLR and the BLR in AGN is set by the incident radiation 
pressure \citep[e.g.][]{paperI, paperII}. Below we describe the 
density and temperature structure induced by RPC in a given gas cloud, and the
resulting free-free emission distribution within the cloud.

\begin{figure*}
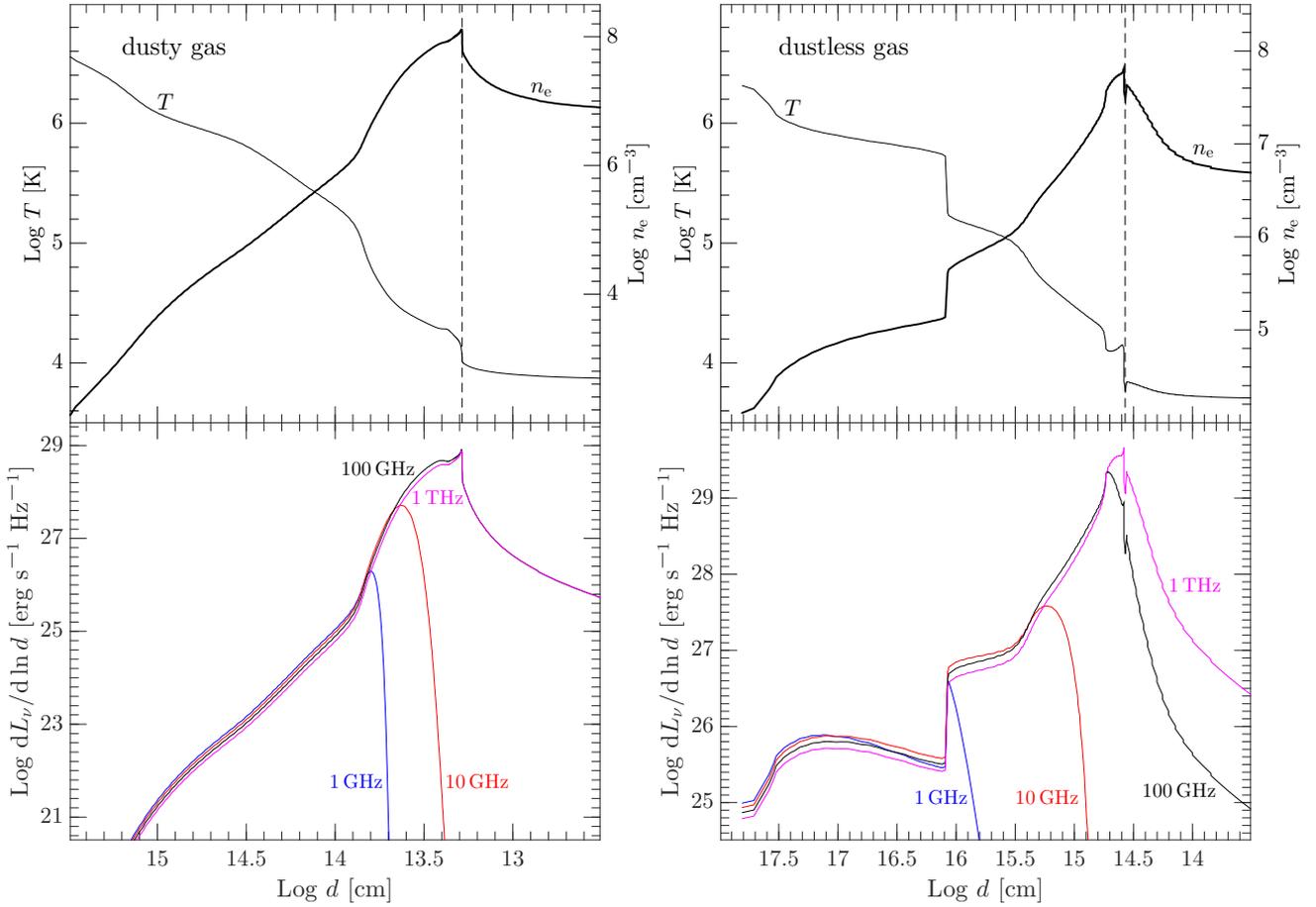

	\includegraphics[width=\columnwidth,trim={0.1cm 0 0 0.01cm},clip]{NLR_Z2_front_emission_r10}
	\includegraphics[width=\columnwidth,trim={0.1cm 0 0 0.01cm},clip]{dustless_NLR_Z2_front_emission_r10}
	\caption{{\bf Upper panels:} 
The electron density and temperature structure of an RPC dusty (left) or dustless (right) slab located at $r=10\rdust$. The position of the ionization front (where $n_e=0.5n$) 
is marked (dashed vertical line). The value of $T$ drops from the Compton value close to the
face of the slab, to $T\approx 10^4$~K at the H ionization front. The value of $n_e$ increases sharply
from the low boundary values to the ionization front, where $n_e\approx 10^8$~cm$^{-3}$. 
The thickness of the dusty gas layer is $\sim 100$ times lower than the thickness of the dustless gas
layer, due to the higher dust opacity.
{\bf Lower panels:} The free-free luminosity per logarithmic unit depth as a function of depth, for several values of $\nu$. 
The calculated emission assumes $\Omega=0.3$ and $L_{46}=1$ (scales linearly with both). 
The emission follows closely the distribution of $n_e$, with a cutoff at a depth where $\tau_\nu> 1$,
and the emission is self-absorbed.
At a high enough frequency, the photoionized layer remains optically thin, and most of
the observed free-free originates close to the ionization front where $n_e$ peaks.}
	\label{fig:RPC_struct_and_em}
\end{figure*}

Figure~\ref{fig:RPC_struct_and_em} (upper panels) presents the temperature and the electron density structure as a function of depth $d$ in the slab, where $d$ is measured from the back side where the H ionization fraction 
drops to 1 per cent (see Section~\ref{sec:numeric_sol}).
We use $d$ rather than the depth from the illuminated face of the slab, since the RPC structure solution is independent of the boundary condition at the illuminated face, when plotted as a function of $d$, i.e.\ the depth from a layer where the ionization fraction drops to a given value (chosen here as 0.01; e.g.\ \citealt{paperII}, fig.~2).
The structure is presented for dusty and dustless gas situated at $r/\rdust = 10$. In both cases, the temperature decreases from $T\gtrsim 10^6$~K near the illuminated side of the slab (the 
Compton temperature; see \citealt{paperII}) to $T\approx 10^4$~K at the H ionization front. 
Beyond the ionization front the gas becomes partly neutral, 
and $T\lesssim10^4$~K. The electron density increases from its assumed initial value at the illuminated side ($\log n_{e,0}= 2$ and 4 for the dusty and dustless model, respectively; Section~\ref{sec:numeric_sol}), to $\log n_e\approx 8$ reached at the H ionization front for both models. 
Inward of the H ionization front, $n_e$ decreases as the gas becomes only partly ionized. The maximal $n_e$ is similar in both models, as it is set by the incident radiation pressure, which is identical for both (eq.~\ref{eq:n}). In the dusty gas, the increase in $n_e$ with decreasing $d$ is smooth (Fig.~\ref{fig:RPC_struct_and_em}, top-left panel), as the increase in the gas pressure is set by
the absorbed radiative flux, absorption set by the dust opacity which is independent of the gas $T$ and $n$. 
For the  dustless gas. the density structure is set by the gas opacity, which depends 
critically on the local $T$ and $n$, leading to sharp changes in the density with $d$ (Fig.~\ref{fig:RPC_struct_and_em}, top-right panel). The compression length scale near the ionization front is $\sim$1~dex smaller for the dusty model since $\sigma_{-21}=1$ for dusty gas, while $\sigma_{-21}=0.1$ for dustless gas near the ionization front, where $U\sim0.1$ (see also eq.~\ref{eq:l_scale}). The dustless gas compression
length scale increases towards the face of the slab, as $U$ increases until the gas is fully ionized,
with only Thomson cross section $\sigma_{-21}=6.65\times10^{-4}$. As a result, the total thickness of the 
dustless slab is larger by $\sim$2~dex than for a dusty slab. 

Figure~\ref{fig:RPC_struct_and_em} (lower panels) presents the free-free luminosity that originates at a given $d$ and emerges from the illuminated face of the slab, at several values of $\nu$. For the purpose of clarity, we plot the luminosity per logarithmic unit depth, 
since the depth extends over $3-4$ orders of magnitude. The emitted luminosity is corrected for the free-free absorption between the emission point $d$ and the illuminated face (eq.~\ref{eq:Lnu_z_esc}). The calculation assumes $\Omega=0.3$ and $L_{46}=1$ (the free-free luminosity scales linearly with $\Omega$ and \Lbol). The local emission follows the density structure since $\epsilon_\nu \sim n_e^2$ (eq.~\ref{eq:em1}), 
as long as the layer is optically thin. Since $\alpha_{\nu} \sim n_e^2 T^{-3/2} \nu^{-2}$ (eq.~\ref{eq:abs_co_approx}), there is a sharp increase in $\tau_\nu$ inwards as $n_e$ increases 
and $T$ decreases. Once $\tau_\nu> 1$ the emission becomes self-absorbed, leading to 
a cutoff in the outgoing emission with decreasing $d$ (Fig.~\ref{fig:RPC_struct_and_em}, bottom panels). 
With increasing $\nu$ the depth (from the illuminated side) where $\tau_\nu> 1$ increases. 
At a large enough $\nu$, the whole photoionized layer becomes optically thin, i.e. $\tau_\nu< 1$
everywhere, leading to 
optically thin free-free emission at that $\nu$. The contribution to the free-free emission from the partly neutral
gas beyond the ionization front decreases due to the decreasing $n_e$. Most of the emission in the
optically thin case, originates from gas close to the ionization front. In the optically thick case,
most of the free-free emission originates at the $\tau_\nu\sim 1$ layer.

\subsection{Free-free absorption} \label{sec:ff_abs}
	
\begin{figure}
		\includegraphics[width=\columnwidth]{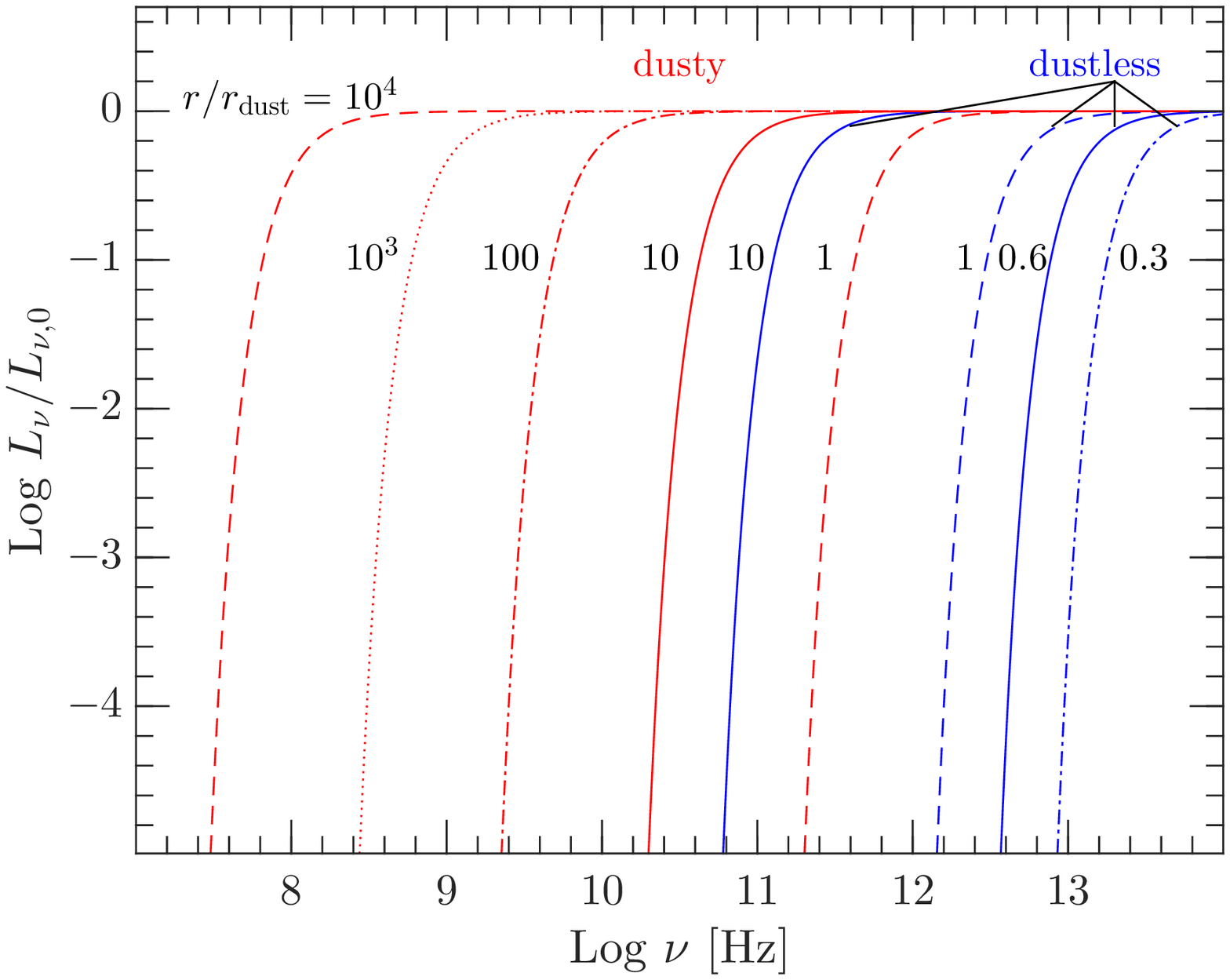}
		\caption{Free-free absorption by dusty and dustless RPC gas located at various distances. The y-axis presents the transmitted luminosity, $L_\nu$, relative to 
the incident luminosity, $L_{\nu,0}$.  The distance is measured in units of \rdust. 
The gas becomes optically thick below the frequency $\nu_{\rm thick}$, which decreases
with increasing distance. At a given distance, $\nu_{\rm thick}$ of dustless gas is higher than
for dusty gas, as the ionized column of dustless gas is larger. 
Dustless gas at the BLR ($r/\rdust< 1$)
becomes optically thick only at $\nu_{\rm thick}< 3\times 10^{12}$~Hz
(or $\lambda> 100$~$\mu$m). Dusty gas at the BLR ($r/\rdust=1$) 
becomes thick  below 500~GHz,
while on the NLR scale ($r/\rdust=10^3$) it remains optically thin down to 1~GHz.}
		\label{fig:abs_only}
	\end{figure}

Figure~\ref{fig:abs_only} presents the free-free absorption of
radiation transmitted through a photoionized RPC gas slab. The slab is located at various distances from the photoionizing source, and is either dusty or dustless. Since the free-free absorption
coefficient $\alpha_\nu \propto \nu^{-2}$ (eq.~\ref{eq:abs_co_approx}) all slabs become optically thick below a certain frequency, $\nu_{\rm{thick}}$. Dustless gas has a higher $\nu_{\rm{thick}}$ than dusty gas at a given location (see Fig.~\ref{fig:abs_only}). The dustless gas has a larger $\nu_{\rm{thick}}$, as the ionized layer extends deeper, i.e.
larger $l$, due to the lower value of $\bar{\sigma}$ (Section \ref{sec:model}), which produces a larger $\tau_\nu=\alpha_\nu l$. 

Photoionized gas close to the BLR is highly absorptive in the radio, up to the mm regime, 
and becomes transparent only in the sub mm for dusty gas ($\nu>300$~GHz, $\lambda<1$~mm),
and only in the FIR for dustless gas ($\nu>3$~THz, $\lambda<100~\mu$m).
In contrast, photoionized gas at the NLR
($r/\rdust>10^3$) becomes transparent already at $\sim 1$~GHz. 

The simple analytic estimate made above (eq.~\ref{eq:v_thin_cases})
gives $\nu_{\rm{thick}}\propto r^{-1}$. The best fit relation for 
$\nu_{\rm{thick}}(r)$, using the Cloudy 
RPC model solutions (Fig.~\ref{fig:abs_only}), yields
	 \begin{equation}
	 \nu_{\rm thick} =
	 \begin{cases}
	  6.03\times10^{11} \left(r/r_{\rm dust}\right)^{-0.95}\text{ Hz}  
& \text{for dusty gas,} \\
	  5.01\times10^{12} \left(r/r_{\rm dust}\right)^{-1.42} \text{ Hz}     
& \text{for dustless gas, } 
	 \end{cases}
	 \label{eq:v_thick_RPC}
	\end{equation}
based on dusty gas solutions at $1\le r/r_{\rm dust}\le 10^4$, 
and dustless gas solutions at $0.3\le r/r_{\rm dust}\le 10$.

The clear spectral signature of free-free absorption is through the presence of a sharp cutoff 
in the observed spectrum below a given frequency, which follows
$\tau_\nu \propto \nu^{-2}$ (eq. \ref{eq:abs_co_approx}). The value of 
$\nu_{\rm{thick}}$ allows to measure directly the distance of the free-free absorbing screen from the ionizing source. 

Free-free absorption of a background radio source can be produced by any foreground ionized gas 
source. The absorbing gas can be heated by other processes, such as
shocks, and not necessarily photoionization. If the distance of the absorbing medium from the central
source is known through other observations, then eq.~\ref{eq:v_thick_RPC} can be used to test
if the absorbing gas is heated by photoionization, and if its density structure is set by RPC. 

If the radio source happens to be free-free emission of photoionized gas, 
then the source free-free emission is always expected to become self-absorbed below $\nu_{\rm{thick}}$, even in the absence of a foreground screen, as further discussed below.

\subsection{Free-free emission}

\subsubsection{Emission from the illuminated side}	
\begin{figure}
		\includegraphics[width=\columnwidth]{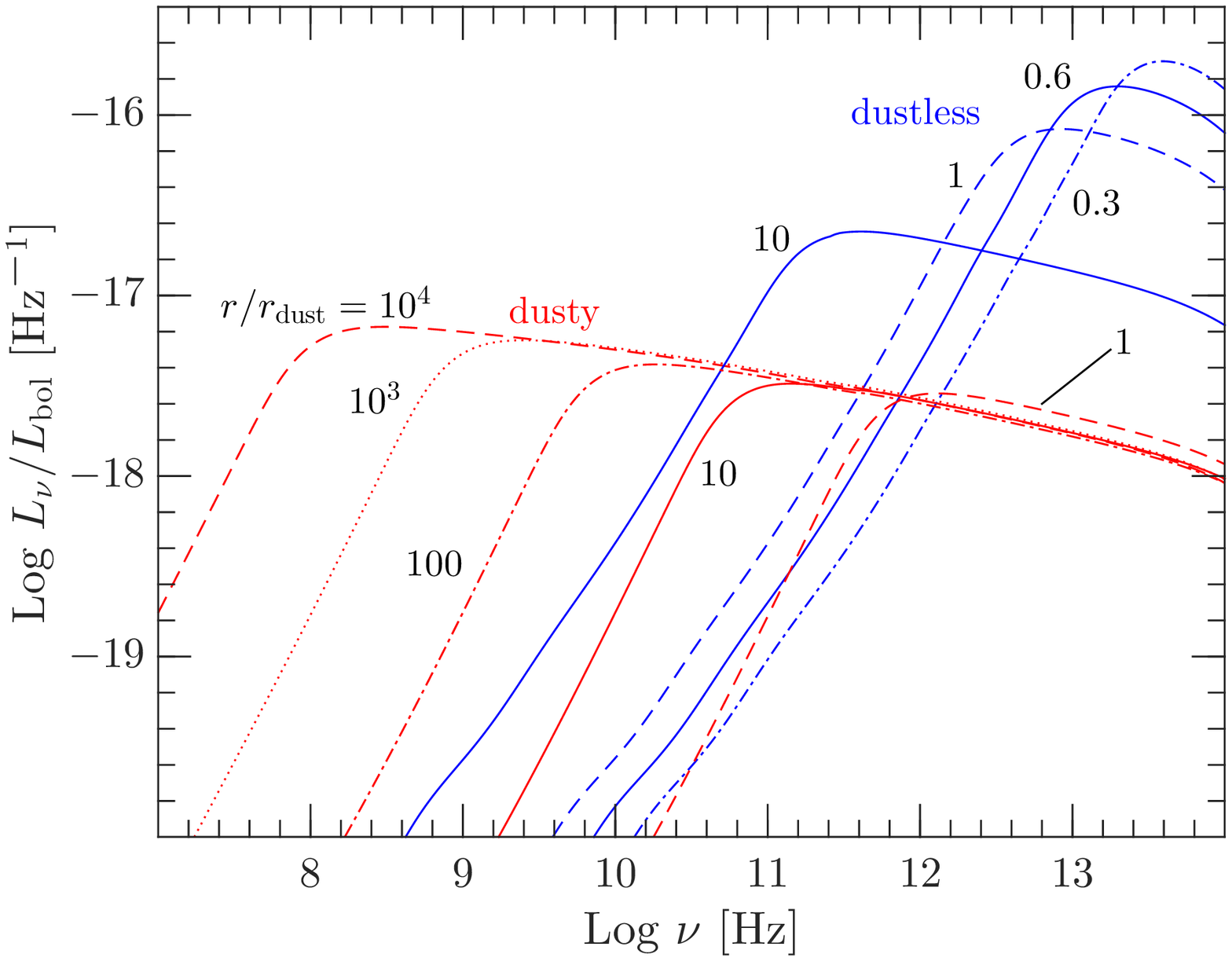}
		\caption{The free-free emission from the illuminated side of a slab located at
various distances. The assumed covering factor is $\Omega=0.3$ at all $r$. The free-free emission spectral slope turns over
from optically thin, $L_\nu \propto \nu^{-0.1}$, to optically thick, 
$L_\nu \propto \nu^{1.4 - 1.64}$ (see text), at $\nu<\nu_{\rm thick}$ (see Fig.~\ref{fig:abs_only}). The optically thin emission of dustless gas
is significantly larger than in dusty gas. Photoionized dustless gas just outside the BLR 
can contribute significantly only at $\nu> 100$~GHz, while optically thin free-free emission at 1~GHz 
can come from photoionized gas only
at $r/\rdust>1000$.}
		\label{fig:ill_side_nu}
	\end{figure}

Figure~\ref{fig:ill_side_nu} presents the emitted spectrum from the illuminated face of a gas slab (eq.~\ref{eq:Lnu_calc}), located at various values of $r/\rdust$. 
The relative strength of the free-free luminosity, $L_\nu/L_{\rm bol}$, is set by 
the covering factor of the slab that is assumed here $\Omega=0.3$, which is typical for the
BLR \citep[e.g.][]{paperII}. 
 At a given distance, the emission at $\nu>\nu_{\rm thick}$ is
optically thin, with $L_\nu \propto \nu^{-0.1}$, and steepens slightly
to $L_\nu \propto \nu^{-0.2}$ above 100~GHz (due to the change in the frequency dependence of 
$\bar{g}_{\rm ff}$). The emission becomes optically thick at $\nu<\nu_{\rm thick}$, and for
an isothermal gas the free-free emission becomes a blackbody, which at the Rayleigh-Jeans regime
gives $L_\nu \propto \nu^2$. For the RPC slab the optically thick spectral slope is somewhat flatter, with
$L_\nu \propto \nu^{1.64}$ for dusty gas, and $L_\nu \propto \nu^{1.4}$ for dustless gas. 
This occurs because the slab is not isothermal, and is composed of
different layers at different temperatures, each one with a different $\nu_{\rm thick}$, leading
to a total emission which is flatter. The spectral turnover at $\nu<\nu_{\rm thick}$ moves to
lower frequencies as $r/\rdust$ increases (Fig.~\ref{fig:abs_only}).

The value of $L_\nu$ in the optically-thin regime of dustless gas decreases with increasing $r/\rdust$, in contrast with the dusty models, where $L_\nu$ remains nearly constant. This difference is due to the following. The emitted free-free spectrum is a function of $T$ (eqs.~\ref{eq:em} and \ref{eq:abs_co}). In dusty gas, the RPC slab structure is mostly set by the dust opacity, which we assume to be independent of $T$ and $n$. Thus, the $T$ structure of a dusty RPC slab is roughly independent of $r/\rdust$. In dustless gas, the structure is set by the gas opacity which depends on $T$ and $n$. Thus, the $T$ structure of a slab varies with $r/\rdust$ (see fig.~3 in \citealt{paperII}). The structure results in a larger value of $L_\nu \sim n^2 r^2 l$, for a given $T\lesssim10^5$~K, with decreasing $r/\rdust$, and thus the maximum $L_\nu$ decreases with $r/\rdust$ for dustless gas. At a given distance, a dustless slab has a larger $L_\nu$ compared to a dusty slab by $\sim1$~dex (see Fig.~\ref{fig:ill_side_nu}). As noted above, the higher emission is a result of a larger compression length-scale $l$ of a dustless slab (Section~\ref{sec:model}).

The free-free luminosity of dustless gas is significantly larger than in dusty gas.
However, dustless gas is expected mostly inside the BLR ($r/\rdust<1$), and since there 
$\nu_{\rm thick}>10^{12}$~Hz, the free-free
contribution of dustless gas in the radio regime is likely negligible. 

Free-free emission at $\nu<1$~GHz must come from gas located at $r/\rdust > 10^3$, where the gas 
is likely to be dusty. Free-free emission at $\nu>100$~GHz can come from gas
located on scales as small as $r/\rdust \sim 10$, which may be dustless (e.g. a failed
disc wind). Thus, the $\nu>100$~GHz regime is where free-free emission from dustless gas is
most likely to be detectable. 

A superposition of free-free emitting gas clouds extending over a range of distances, 
say from $r_{\rm in}$ to $r_{\rm out}$, can produce
a spectral slope, $\alpha$, at a given $\nu$ which can be anywhere in the range  
$\alpha_{\rm thin} <\alpha < \alpha_{\rm thick}$, or specifically $-0.1<\alpha<1.64$,
for $\nu$ in the intermediate range  $\nu_{\rm thick}(r_{\rm out})<\nu<\nu_{\rm thick}(r_{\rm in})$. 
At $\nu>\nu_{\rm thick}(r_{\rm in})$ all contributing gas clouds become optically
thin, which gives $\alpha \approx -0.1$, while at $\nu<\nu_{\rm thick}(r_{\rm out})$
all clouds are optically thick, and $L_{\nu}$ will fall as fast as $\alpha \simeq 1.64$
(dusty gas) or $\alpha \simeq 1.4$ (dustless gas). 
At the intermediate $\nu$ values, the emission at a given $\nu$
is dominated by the clouds where $\tau_\nu\sim 1$. This allows to measure directly
the emitting surface area of the free-free emitting clouds, based only on the value
of $L_{\nu}$, as further described below.

\begin{figure}
		\includegraphics[width=\columnwidth]{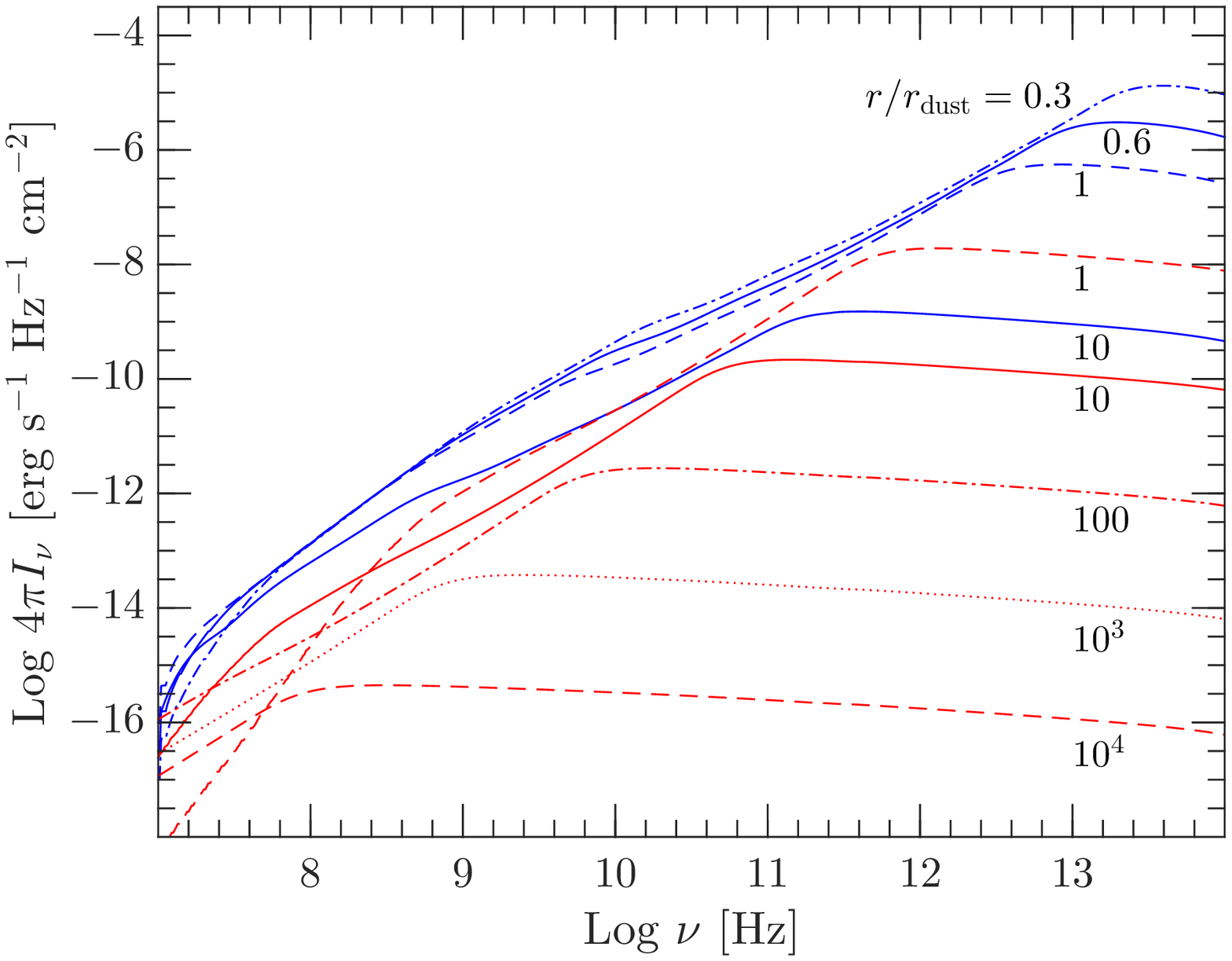}
		\caption{The same as Fig.~\ref{fig:ill_side_nu} for the specific intensity $I_\nu$ --
the flux per solid angle subtended by the emitting gas.
The optically thick $I_\nu$ approaches the blackbody limit, at the temperature
of the free-free emitting region. This maximal intensity, at a given $\nu$, 
allows to derive the minimal emitting surface area from the observed $L_{\nu}$. 
If the gas is optically thin, as implied by the observed
spectral slope, the minimal emitting area is larger.}
		\label{fig:ill_side_nu_rel_r2}
	\end{figure}

	Figure~\ref{fig:ill_side_nu_rel_r2} presents the specific intensity, $I_\nu$, multiplied by $4\pi$, 
i.e. the flux density emitted by a unit area, assuming isotropic emission. 
The radiation is produced by the slabs presented in Fig.~\ref{fig:ill_side_nu}. 
The solutions
for dustless BLR slabs ($r/\rdust\leq1$) at $\nu<\nu_{\rm thick}$ all collapse to a nearly single solution, which 
corresponds to the quasi-blackbody emission produced in slabs with similar temperatures. The small range of $T$ results from the limited range of $0.3 \lesssim r/\rdust\leq 1$ that corresponds to the BLR. The value of $I_\nu$ is independent of distance
from the central source, and is also independent of $L_{\rm bol}$. 
At $\nu>\nu_{\rm thick}$ the layer becomes optically thin, and $I_\nu$ drops below the maximal 
blackbody value obtained at $\nu<\nu_{\rm thick}$. 
The value of the predicted $I_\nu$ allows to constrain the minimal possible emitting surface area 
$A$, since $L_\nu=A\times \pi I_\nu$. 

For dusty gas, which extends over a broader range of values of $r/\rdust$, we get a similar behaviour, although $I_\nu$ of the optically-thick gas decreases slightly
with increasing distance. This occurs due to the following. The optical depth at a given $\nu$ is approximately $\tau_\nu \propto T^{-5/2} (r/\rdust)^{-2}$ for $h\nu\ll kT$ (eq.~\ref{eq:abs_co}), where we use $n\propto T^{-1} (r/\rdust)^{-2}$ (\citealt{paperI}, eq.~6) and $l\propto T (r/\rdust)^2$ (eq.~\ref{eq:l_scale}), and assume that $\bar{\sigma}$ is independent of $T$ and $n$.
Thus, the maximal temperature of a layer which is still optically thick at $\nu$ (i.e.\ $\tau_\nu\approx1$) decreases with $r/\rdust$, yielding the decrease of $I_\nu$ with $r/\rdust$ of the optically-thick gas.
 
Note the spectral break of the $r/\rdust=1$ gas emission below 1~GHz. This break occurs since $\nu<10\nu_{\rm p}$ where we assume $\epsilon_\nu=0$ (Section~\ref{sec:numeric_sol}).

If the free-free emitting source is spatially resolved, the value of $I_\nu$ can be directly determined.
The value of $I_\nu$ is conveniently expressed in the radio regime using the brightness temperature, 
$T_{\rm b}\equiv I_\nu c^2/2\nu^2 k$. An isothermal gas slab at a temperature $T_{\rm e}$ 
with an optical depth $\tau_\nu$, produces emission with $T_{\rm b}=T_{\rm e}(1-e^{-\tau_\nu})$
(e.g.\ \citealt{RL04}). So, for optically thick gas we get blackbody emission, with $T_{\rm b}=T_{\rm e}$,
and for optically thin gas we get $T_{\rm b}=T_{\rm e}\tau_\nu$.
Since RPC gas is not isothermal, the derivation of $T_{\rm b}$ is somewhat less trivial.

\begin{figure}
		\includegraphics[width=\columnwidth]{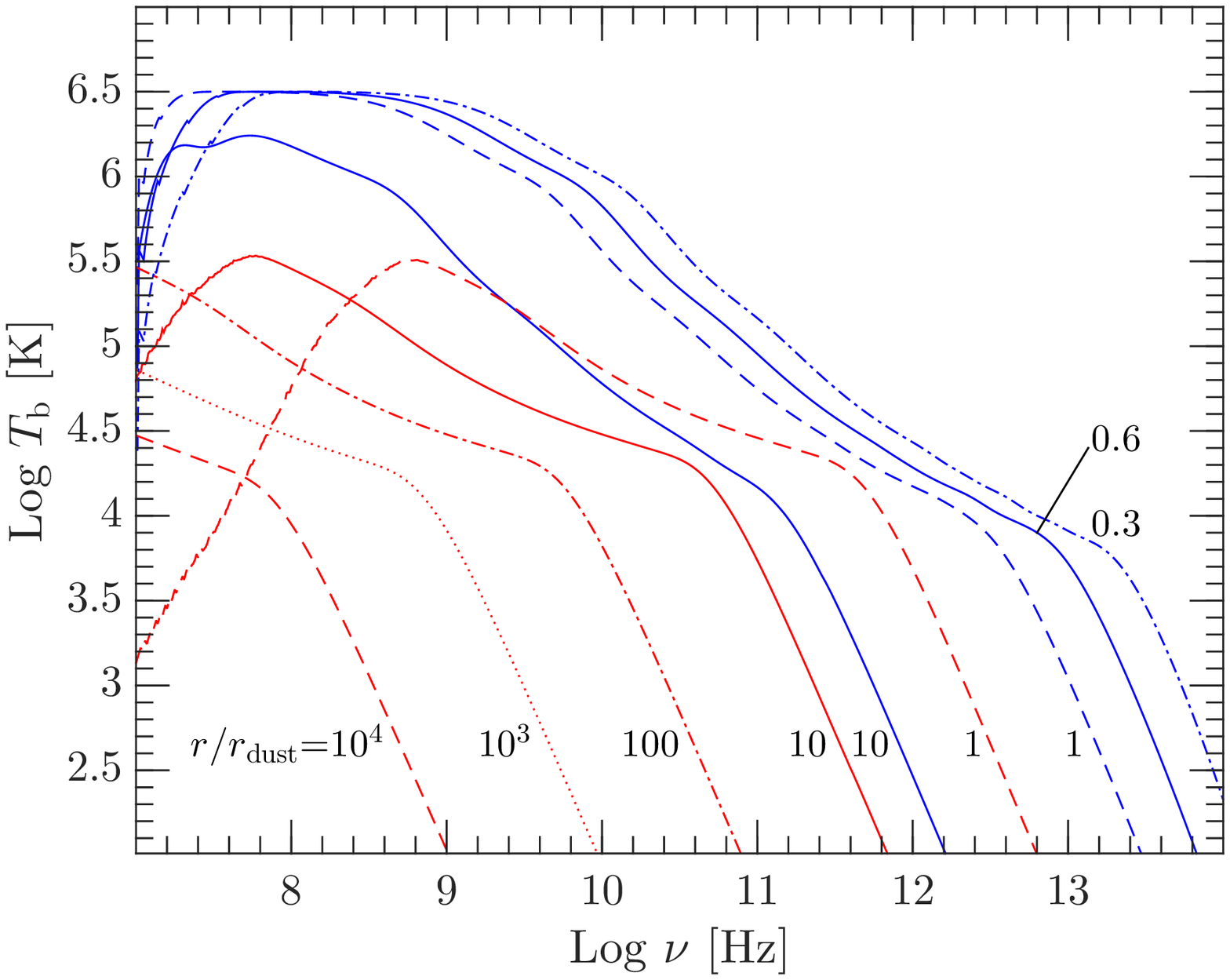}
		\caption{Same as 
Fig.~\ref{fig:ill_side_nu_rel_r2} where $I_\nu$ is replaced 
by $T_{\rm b}=I_\nu c^2/2\nu^2 k$, a convenient way to represent $I_\nu$. 
At $\nu>\nu_{\rm thick}$
(see values in Fig.~\ref{fig:abs_only}), an RPC slab is optically thin, which gives 
$T_{\rm b}\propto \nu^{-2}$, as $I_\nu\sim $ constant. 
At $\nu<\nu_{\rm thick}$, the slab is optically thick, and
$T_{\rm b}\sim T_{\rm e}$ at the $\tau_\nu\sim 1$ layer. With decreasing $\nu$, 
$\alpha_\nu$ increases, and thus the $\tau_\nu\sim 1$ layer is found closer to the surface, 
where the gas is less compressed and $T_{\rm e}$ is higher. The steep drop at the lowest
$\nu$ is due to self-absorption below the plasma frequency (eq.\ref{eq:plasma}). In all models
$T_{\rm b}\le 3\times 10^6$~K, the Compton temperature of the gas. On scales of
$r/\rdust>10$, which can be resolved in nearby AGN at $\nu>10$~GHz, the maximal 
$T_{\rm b}$ of free-free emission is $\le 10^5$~K (see also Fig.~\ref{fig:const_n_Tb}). 
}
		\label{fig:Tb}
	\end{figure}

Figure~\ref{fig:Tb} presents $T_{\rm b}$ for the same
set of RPC models presented in Fig.~\ref{fig:ill_side_nu_rel_r2}. A simple and robust conclusion 
is that a source with $T_{\rm b}>3\times 10^6$~K cannot be produced by free-free emission from RPC gas. 
Lower maximal values for $T_{\rm b}$ are possible at $\nu>1$~GHz. Dusty gas always has a lower
$T_{\rm b}$ than dustless gas, as the ionized column is about a factor of 10 smaller than in dustless
gas ($\sim$90 per cent of the ionizing continuum is absorbed by the dust and is reradiated in the IR). 
The steep drop in $T_{\rm b}$ occurs when $\tau_\nu< 1$, as $T_{\rm b}\propto \tau_\nu$. 
A shallower drop of $T_{\rm b}$ occurs in the regime where
$\tau_\nu\gg 1$. The drop in $T_{\rm b}$ with increasing $\nu$ occurs as the layer
where the $\tau_\nu\sim 1$, which dominates the emission, is thicker and extends inwards to the colder deeper layers where the gas
is denser and less ionized.

\subsubsection{Back side emission}	
	
\begin{figure}
		\includegraphics[width=\columnwidth]{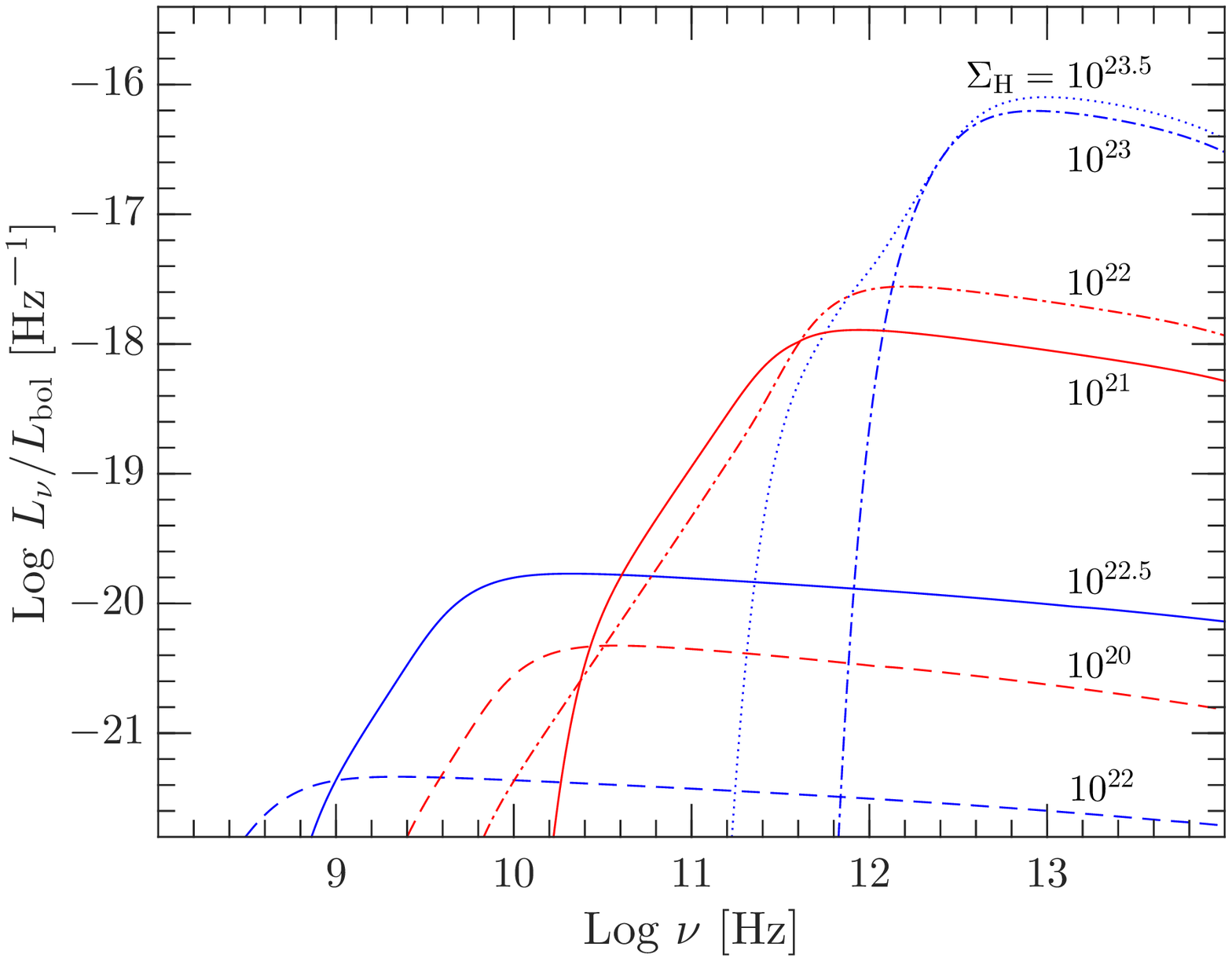}
		\caption{
The free-free emission from the back side of a dusty (red) or a dustless (blue) slab, 
as a function of the slab total H column \NH. 
The slab is located at $r/\rdust=1$. The optically thin emission 
increases up to a column of \NH$\sim 10^{23}$~\cmmt\ for dustless gas, and
$\sim 10^{21}$~\cmmt\ for dusty gas. Above these columns the additional gas resides 
behind the ionization front, the gas is 
mostly neutral, and produces little free-free emission or absorption. 
Most of the emission is produced close to the ionization front 
since $\epsilon_\nu \propto n_e^2T^{-1/2}$ (eq.~\ref{eq:em}),
and $n_e$ is highest there, and $T$ is lower. 
At lower \NH, $L_{\nu}$ drops sharply, by a factor of $\sim  10^5$ 
at \NH$\sim 10^{22}$~\cmmt\ in dustless gas, and by $\sim 300$ at
 \NH$\sim 10^{20}$~\cmmt\ for dusty gas. Also, $\nu_{\rm thick}$ decreases steeply 
with decreasing \NH, by a factor of $\sim 3000$ for dustless gas, and by
 $\sim 30$ for dusty gas, at the above columns.}
		\label{fig:em_vs_NH}
	\end{figure}

	Figure~\ref{fig:em_vs_NH} presents the emitted spectrum from the back side of a dusty or dustless RPC slab, for a range of values of a total H column \NH. The slab is located at $r/\rdust=1$ in all cases. The maximal values of $\NH=10^{23.5}$ and $10^{22}$~\cmmt\ are presented for a dustless and a dusty slabs. Increasing \NH\ further has little effect on the emitted spectrum, since the gas  becomes mostly neutral, and does not contribute to the emission or to the absorption
($\epsilon_\nu\approx0$, $\alpha_\nu\approx 0$). 

The value of $L_\nu$ in the optically thin regime 
increases with \NH. In the dustless gas the effect can be dramatic,
where an increase of \NH\ from $10^{22}$ to $10^{23}$~\cmmt\ yields an increase of 
$L_\nu$ by $\sim$5~dex. A further increase from $\NH=10^{23}$ to $10^{23.5}$~\cmmt\ yields
only a small rise, as the additional column resides behind the ionization front, and is mostly neutral. A similar, but less dramatic, 
effect is found for the dusty slab, where $L_\nu$ increases by $\sim$2~dex when 
\NH\ increases from $10^{20}$ to $10^{21}$~\cmmt, with only a small rise when
\NH\ increases further to $10^{22}$~\cmmt. 

The sharp increase in $L_\nu$ in the optically thin dustless gas occurs since $\epsilon_\nu\propto n^2$ (eq.~\ref{eq:em}), and $n$ rises sharply near the ionization front in dustless gas due to the
RPC effect (e.g.\ \citealt{paperII}, fig.~3). In dusty gas, the UV opacity is
independent of the gas ionization, and the rise in $n$, due to the RPC effect, is less
sharp (see \citealt{paperI}, fig.~2; note also that $\tau\propto\NH$ for dusty gas).

Fig.~\ref{fig:em_vs_NH} also shows that $\nu_{\rm thick}$ increases with \NH. 
This occurs since $n$ increases with depth, i.e.\ \NH, as the gas compression 
becomes larger inwards in the RPC solution.
Thus, as $\nu_{\rm thick}$ is set by $\tau_{\nu} = \alpha_{\nu}l =1$, and  $\alpha_{\nu}\propto n^2\nu^{-2}$ (eq.~\ref{eq:abs_co_approx}), $\nu_{\rm thick}$ increases (the decrease in $l$ with \NH\ is typically smaller than the increase in $n$; e.g.\ \citealt{paperII}, figs~1 and 2).
 The drop in $L_\nu$
at $\nu<\nu_{\rm thick}$ is very sharp, in particular for the dustless gas. This occurs since 
the observed radiation comes only from the $\tau_{\nu} \sim 1$ layer, and the thickness of
this layer drops with decreasing $\nu$ as $\alpha_{\nu}^{-1}\propto\nu^2$.
At $\nu>\nu_{\rm thick}$, we get $\tau_{\nu} <1$ throughout the
slab, and the emission is isotropic.
	
	\subsection{Comparison with uniform density  models} \label{sec:compare_const_n}

How do the RPC results presented above differ from the predicted free-free emission 
of the commonly assumed uniform-density gas clouds?
	
\begin{figure}
		\includegraphics[width=\columnwidth]{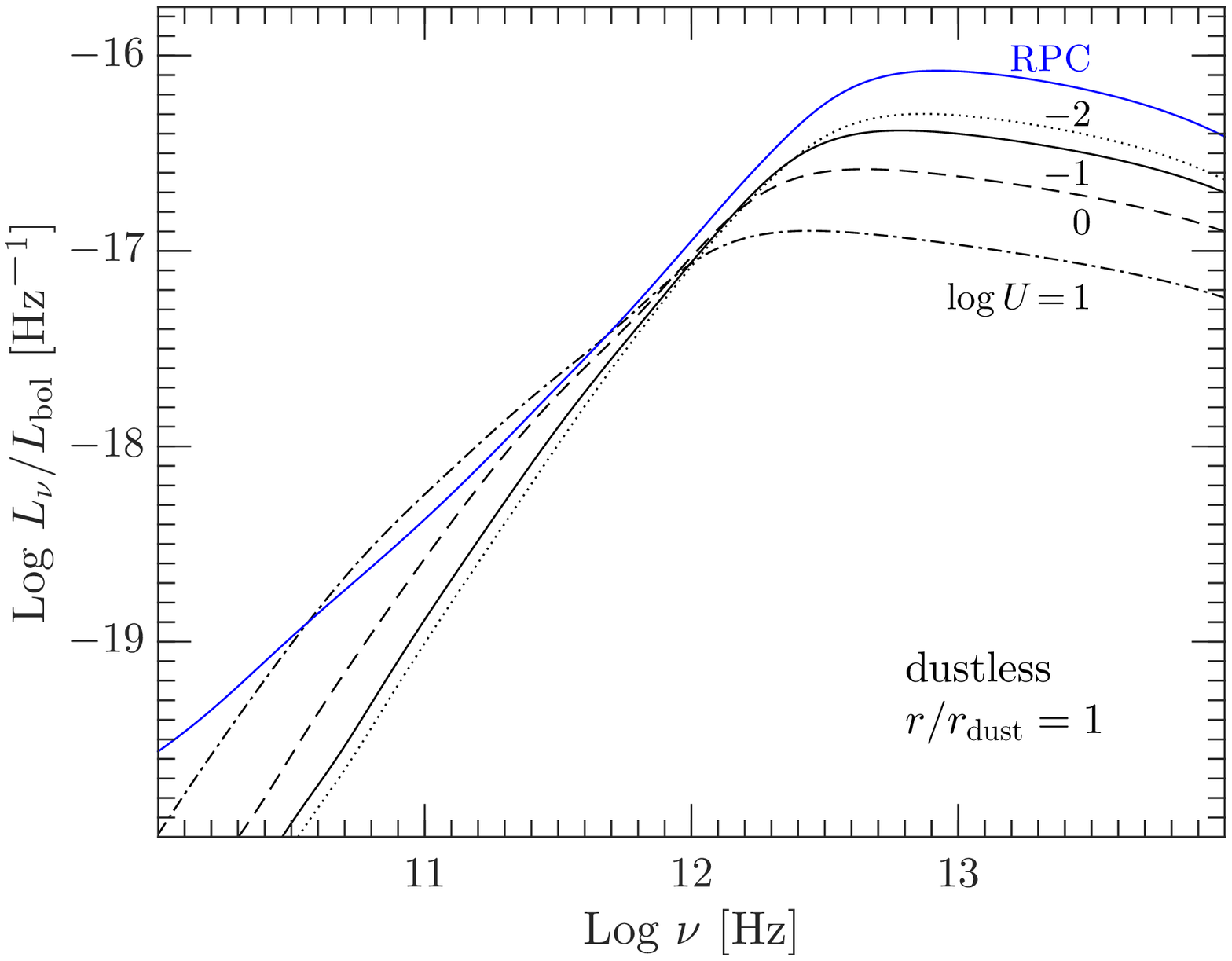}
		\caption{Comparison of the free-free emission of RPC gas with the 
emission of uniform density gas with various $U$ values. The gas is dustless and is situated at $r/\rdust=1$, 
where $\log n = 8.57 - \log U$. 
The emission of a uniform density slab depends on the assumed value of
$U$, in contrast with the RPC solution which is uniquely set by the distance.
The optically thick emission of the uniform density gas generally shows a steeper 
($\alpha\sim 2$) optically thick emission, compared to RPC gas. This occurs since a uniform density slab is closer to being isothermal, compared
to the RPC slab, which is effectively composed of a superposition of layers with 
decreasing $T$ inwards, leading to a flatter
slope ($\alpha \sim 1.64$, Fig.~\ref{fig:ill_side_nu}).
}
		\label{fig:BLR_const_n10}
	\end{figure}

\begin{figure}
		\includegraphics[width=\columnwidth]{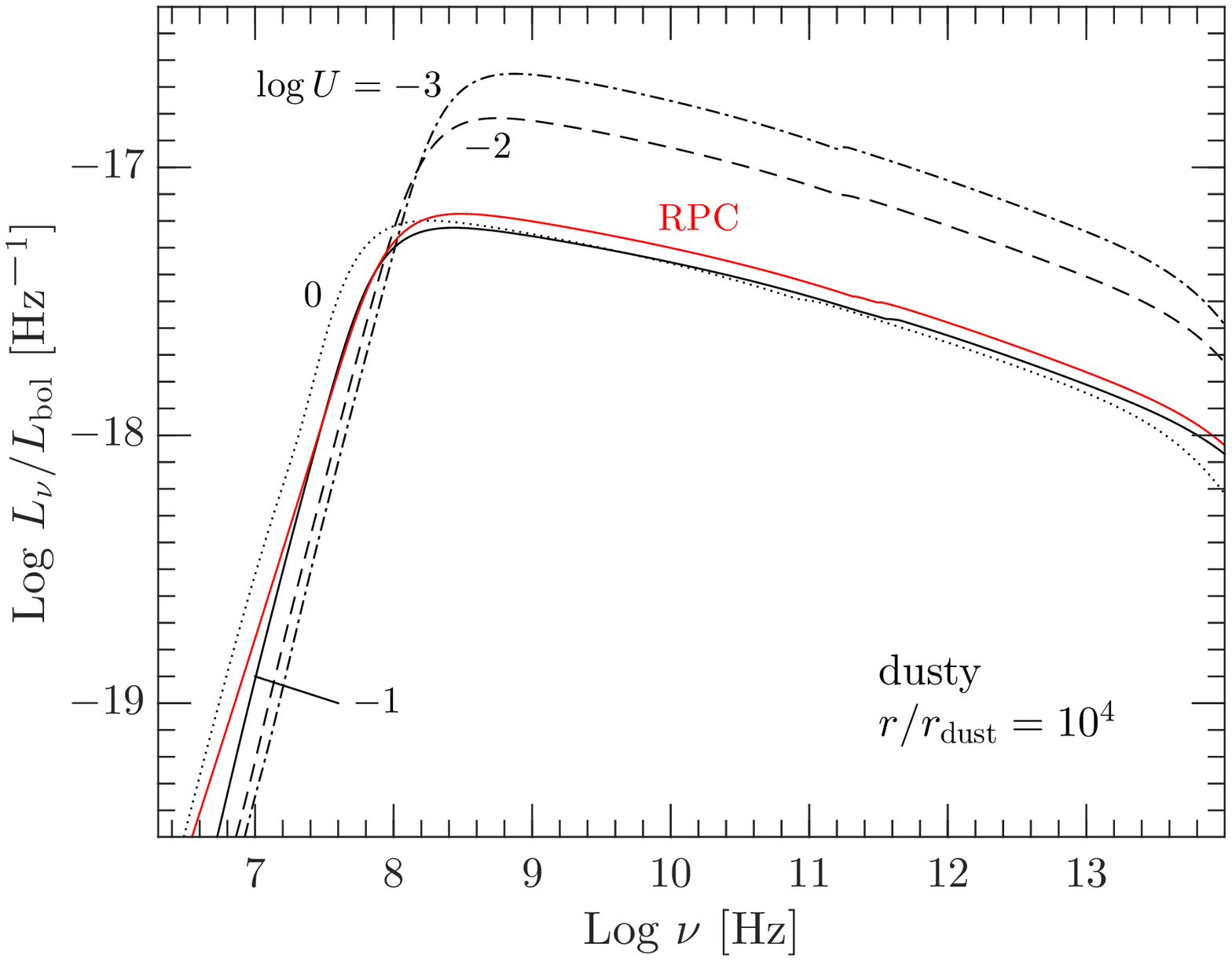}
		\caption{Same as Fig.~\ref{fig:BLR_const_n10} for a dusty
slab at $r/\rdust=10^4$, with various values of $U$, compared to an RPC slab at the same distance. The density is $\log n = 0.57 - \log U$.  The optically thin spectral slope is similar for all
models, and the optically thick RPC spectral slope is only slightly flatter than the slope of the constant density slab.
}
		\label{fig:NLR_const_n3}
	\end{figure}

	Figure~\ref{fig:BLR_const_n10}, and Figure \ref{fig:NLR_const_n3}, compare the calculated free-free emission of uniform-density dustless slab, and dusty slab, respectively, with that of an RPC slab located at the same distance. The emitted spectrum is calculated from the illuminated side. In uniform density models the gas density,
or equivalently the ionization parameter, need to be assumed.
The dustless models (Fig.~\ref{fig:BLR_const_n10}) are evaluated for 
$-3\le \log U\le 1$ at $r/\rdust = 1$. The associated gas density is 
 $\log n = 8.57 - \log U$. The dusty models 
(Fig.~\ref{fig:NLR_const_n3}) are evaluated at $r/\rdust = 10^4$
for $-3\le \log U\le 0$, where the associated density is $\log n = 0.57 - \log U$.

The uniform density models predict emission spectra which depend on the value of $U$.
The optically thin emission decreases with increasing $U$ in both dusty and dustless gas.
This occurs for the following reason. The optically thin emission follows 
$L_\nu \propto \epsilon_\nu l$, where
$\epsilon_\nu \propto T^{-1/2} n^2$ (eq.~\ref{eq:em}). In the dustless slab, 
the column of the ionized layer
satisfies $ln\propto  U$. Thus, $L_\nu \propto T^{-1/2} n U\propto T^{-1/2} n_\gamma$,
or $L_\nu \propto T^{-1/2}$, at a given $r$ where $n_\gamma$ is fixed. The rise in $L_\nu$ with decreasing $U$ in dustless uniform density gas (Fig.~\ref{fig:BLR_const_n10}) therefore occurs because of the drop in $T$.

In dusty gas, the ionized column $ln$ for $\log U> -2$ is set by the dust UV opacity, which is constant.
Thus, $L_\nu \propto T^{-1/2} n \propto T^{-1/2} U^{-1}$, which leads to a steeper drop
in $L_\nu$ with increasing $U$. For $\log U\le -2$ the gas opacity dominates, so the increase
in $L_\nu$ with decreasing $U$ becomes smaller, and is similar to the change in dustless gas.

The transition to optically thick emission occurs at $\nu_{\rm thick}$, where 
$\tau_\nu=\alpha_\nu l=1$. Since $\alpha_\nu\propto T^{-3/2} n^2 \nu^{-2}$
(eq.~\ref{eq:abs_co_approx}), and in dustless gas $ln\propto U$, we get $\tau_\nu\propto T^{-3/2} \nu^{-2}$,
which gives $\nu_{\rm thick}\propto T^{-3/4}$ at the $\tau_\nu\sim 1$ layer. 
Thus, $\nu_{\rm thick}$ decreases with increasing
$U$ since $T$ increases. In dusty gas at $\log U>-2$ the ionized column 
$l n$ is constant, which
gives $\nu_{\rm thick}\propto T^{-3/4}n^{1/2}$, which results in a somewhat steeper drop of 
$\nu_{\rm thick}$ with increasing $U$.

The spectral slope of the optically thick emission of dustless gas at $\log U \leq 0$
is close to 2, the Rayleigh-Jeans slope of blackbody emission. This reflects the small
temperature gradient within a uniform density ionized slab. 
The optically thick slope becomes somewhat flatter for $\log U =1$, reflecting the increase in $T$ close to the
surface where $T \sim 10^5$~K (compared to $T\sim 10^4$~K inwards). The superposition of free-free emission at
different temperatures leads to the spectral flattening. 
In dusty gas, the optically thick emission shows less structure,
as the slab structure
is set by the fixed dust opacity, which leads to a more isothermal structure.
	
	The comparison above highlights the advantage of the RPC solution. Apart from 
being more realistic, the predicted gas emission is set
only by a single free parameter, $r/\rdust$. The distance of the photoionized gas can sometimes be observationally determined (see below), which then leaves no free parameters. 
In contrast, 
the uniform density model predictions depend on an additional free parameter, $n$, or $U$.
The attempts to determine $U$ from line emission observations imply that it spans a broad range of values  \citep{baldwin_etal95}, a range which naturally results from the RPC effect
\citep{paperII, paperIII}.

\begin{figure}
		\includegraphics[width=\columnwidth]{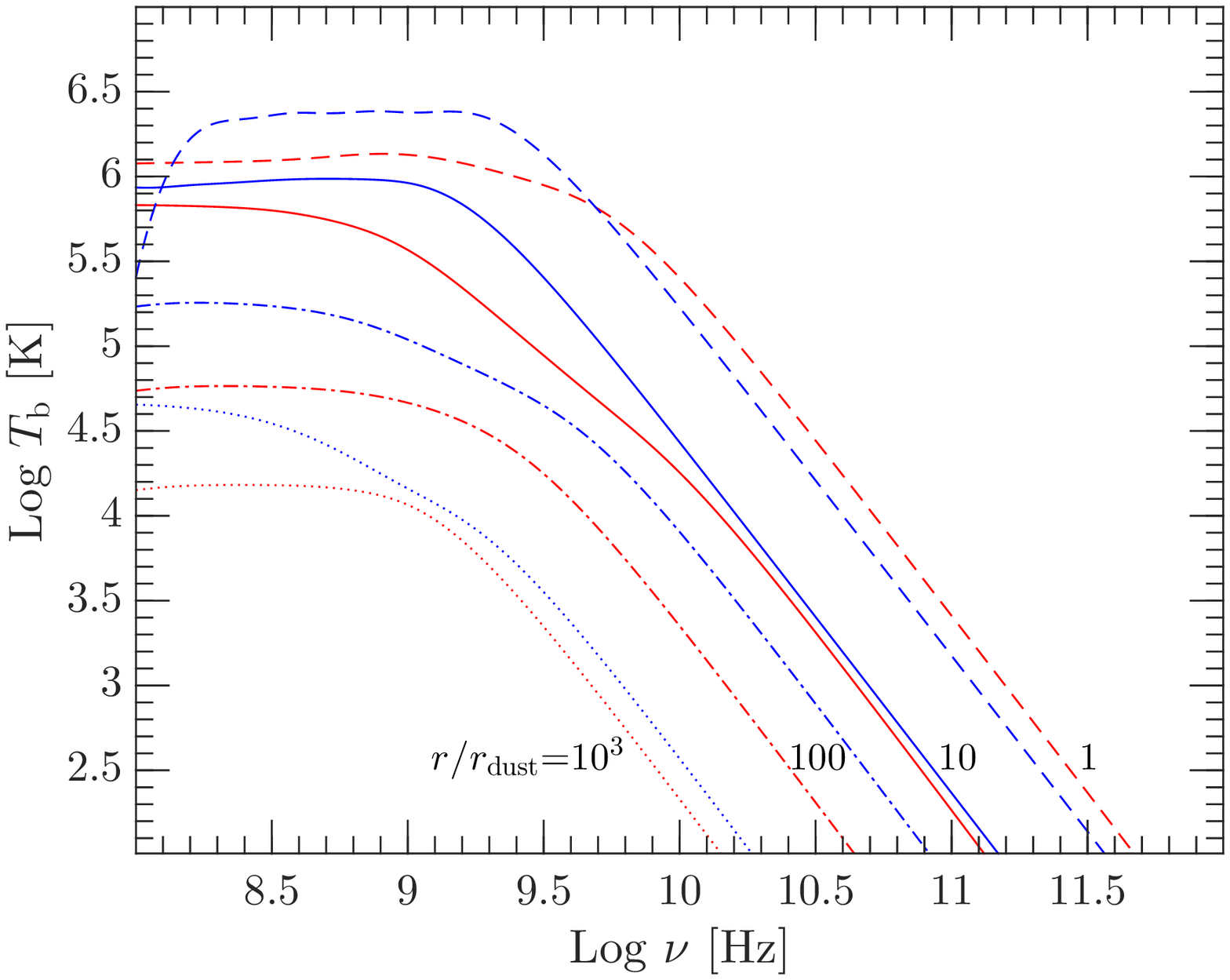}
		\caption{Same as Fig.~\ref{fig:Tb} for a constant density slab. 
At each value of $r/\rdust$ we present the dustless and dusty slabs with the value
of $U$ which leads to the maximal value of $T_{\rm b}$. These are generally the highest
$U$ and highest $\tau_\nu$ slabs which can physically fit within the available $r$. 
These criteria lead to
the highest $T_{\rm e}$ close to the surface (where $\tau_\nu=1$), as required since 
$T_{\rm b}\le T_{\rm e}$. A radio source with $T_{\rm b}> 10^6$~K at 
$\nu>$ a few GHz cannot be produced by free-free emission of photoionized gas, regardless
of the gas confinement mechanism (see also Fig.~\ref{fig:Tb}).
}
		\label{fig:const_n_Tb}
	\end{figure}

Figure~\ref{fig:const_n_Tb} presents the maximal possible brightness temperature 
$T_{\rm b}$ of uniform density slabs at a range of distances. As in the RPC case 
(Fig.~\ref{fig:Tb}), the maximal value is $T_{\rm b}\sim 3\times 10^6$~K, and is
achieved only at $\nu < 1$~GHz, and for extremely compact clouds
at $r\sim \rdust$. Radio emission with
$T_{\rm b}> 10^6$~K at $\nu>$ a few GHz, cannot be generally produced by photoionized
gas, regardless of the gas location or density. In fact, such $T_{\rm b}$ values
cannot be produced by free-free emission of hot gas in RQ AGN in general, regardless 
of the gas heating mechanism, as further discussed below.

\section{Comparison with observations} \label{sec:comare_with_obs}

\subsection{Free-free absorption} \label{sec:obs_ff_abs}

Free-free absorption of a background continuum source 
produces a sharp spectral break below a certain frequency, $\nu_0$, 
with a flux drop of the form $L_\nu\propto e^{-(\nu_0/\nu)^2}$. This is in contrast with 
the spectral break due to self-absorption of either a free-free source,
or a synchrotron source, which produces a power-law (PL) below the spectral break frequency, 
with $L_\nu \propto \nu^2$ or $L_\nu \propto \nu^{2.5}$ \citep[e.g.][]{RL04}.
Thus, free-free absorption has a unique characteristic. In addition, free-free absorption 
can be easy to detect,  if it occurs in front of a high-brightness-temperature
background source, which is often present in radio loud AGN. A sharp spectral break below some 
frequency, suggestive
of free-free absorption, is indeed detected in some radio loud AGN \citep{GK14, Call17, Mh19},
and is sometimes spatially resolved in high resolution observations
\cite[e.g.][]{V94, W94, LLV95, U99,U99a, G99, Ped99, W00, Jones01, Marr01, Kameno03}. Below we discuss a few well studied cases in nearby AGN, and their consistency with the RPC model
predictions for a free-free absorption screen.

Apparently the best studied free-free absorption case is in NGC~1275 (3C~84),
a radio loud active galaxy in the centre of the Perseus cluster. High angular resolution VLBA
observations at 5~GHz showed historically a single sided jet, as commonly observed in highly beamed sources
pointing at us. However, followup VLBA observations at 15 and 22~GHz
revealed a counter jet $6-8$~mas to the north, a feature which is free-free absorbed at lower frequencies \citep{V94, W94}. 
The free-free absorbing screen is likely in a disc configuration, lying in front of the counter jet,
and behind the jet directed at us \citep{V94,W94}. 
Is the free-free absorbing screen composed of photoionized gas? 
If yes, is the optical depth of the screen, which is spatially resolved in this object, consistent with the 
distance versus absorption frequency relation of RPC gas?

The absorption is spatially resolved with a projected distance on the sky of $2-2.5$~pc from the centre 
\citep{V94, W94, LLV95, W00}. 
The jet is inclined at $\theta\sim 65^{\circ}$ to the line of sight \citep{FN17}. If the absorbing
screen is a disc at right angle to the jet, then 
the projected absorber distance corresponds to a physical distance along the disc 
of $4.7-5.9$~pc from the ionizing source.
The bolometric luminosity of NGC~1275 is 
$\Lbol\simeq 4\times 10^{44}$~\ergs\ (section 2.2 in \citealt{LLV95}), which gives
$\rdust=0.04$~pc (eq.~\ref{eq:rdust}). Thus, the absorbing gas resides at a deprojected distance of
$r/\rdust\simeq 120-150$.
The photoionized RPC gas gives $\nu_{\rm thick}\simeq 5.2-6.4$~GHz 
for dusty gas, and $\nu_{\rm thick}\simeq 4.1-5.6$~GHz extrapolated for dustless gas, using 
the relations in eq.~\ref{eq:v_thick_RPC}. 
The observations yield $\tau_{8~{\rm GHz}}\simeq 2$ \citep{LLV95}, which implies $\nu_{\rm thick}\simeq 11$~GHz,
which agrees within a factor of two with the predicted values. 
Thus, the free-free screen observed
in NGC~1275 is consistent with RPC photoionized gas, where both the ionization structure and the density
structures are set by the illuminating ionizing radiation.

A similar spatially resolved free-free absorber is detected in NGC~4151,
where \cite{Ped99} detected free-free absorption with $\nu_{\rm thick}\simeq 300$~MHz
of an extended radio source located $\sim 50$~pc from the centre. 
This AGN shows significant long term variablty, with a typical flux of  
$F_\lambda(5100\text{\AA})=4\times 10^{-14}$~\ergs~cm$^{-2}$~\AA$^{-1}$ measured
20 years prior to the radio observations \citep{Onken07}. This luminosity may not 
represent the luminosity $\sim$150 years ago, which is illuminating now the screen. 
However, this luminosity implies
$\log L_{\rm bol}\simeq 43.6$, and thus $\rdust\simeq 0.013$~pc. The absorber 
therefore extends over a distance of $r/\rdust\sim 3850$. 
The RPC predicts $\nu_{\rm thick}\simeq 240$~MHz, consistent 
with the observed $\nu_{\rm thick}\simeq 300$~MHz. 

Another object where the free-free absorption is spatially resolved is NGC~4261, 
where $\tau_{5~{\rm GHz}}< 1$ at $r>0.5$~pc \citep{Jones01}. The estimated
$\log L_{\rm bol}\simeq 41.8$ \citep{Erac10} implies $\rdust\simeq 1.65\times 10^{-3}$~pc,
and therefore the observations indicate that $\tau_{5~{\rm GHz}}< 1$ at $r/\rdust>300$. 
This result is  consistent with the RPC prediction
that $\tau_{2.6~{\rm GHz}}= 1$ for $r/\rdust= 300$.

The case of NGC~1052 is particularly interesting \citep{Kameno03}, as 
the observations show a clear rise in $\nu_{\rm thick}$ with decreasing distance
from the centre. Specifically,
fig.~3 in \cite{Kameno03} suggests that $\nu_{\rm thick}\sim 3$, $\sim 6$ and 
$\sim 20$~GHz, at projected distances of $\sim 1.7$, $\sim 0.9$ and
$\sim 0.4$~pc from the centre (using the 0.1~pc/mas scaling in this nearby galaxy).
The estimated $\log L_{\rm bol}\simeq 43.15$ (using $L_{\rm bol}=10L_{\rm X}$ and
$L_{\rm X}$ from \citealt{Cus10}), gives $\rdust\simeq 7.5\times 10^{-3}$~pc.
Thus, the projected distances are $r/\rdust\sim 225, 120$ and 53, and the
RPC predicted values for $\nu_{\rm thick}$ are 3.5, 6.4 and 13.9~GHz,
which agree well with the above estimated values of $\sim 3$, 
$\sim 6$ and $\sim 20$~GHz. Clearly, a further study of this object is warranted.

\subsection{Free-free absorption and emission in NGC 1068} \label{sec:ngc1068}
 
Possible detection of free-free emission was made in
NGC 1068, the archetype of obscured AGN. High resolution VLBA observations at $1.4-8.4$~GHz 
revealed a 
compact nuclear source, which extends on a scale of 
$\sim 10$~mas, at right angle to the large scale outflow. The emission spectrum is a
flat PL with $\alpha=-0.17\pm 0.24$ at $5-8.4$~GHz, with $T_{\rm b}\simeq 4\times 10^6$~K \citep{G97,Roy98,G04}. 
This feature was interpreted as free-free
emission from highly photoionized gas which covers the surface of the obscuring torus gas, a layer produced by 
the X-ray illumination of the central ionizing source. However, this PL component cannot be produced
by free-free emission, as further described below.

The compact $\sim$10~mas scale PL source is characterised by $\langle T_{\rm b}\rangle =4\times 10^6$~K at 
5 GHz, with a peak value of $8.6\times 10^6$~K. The emission is interpreted as free-free emission 
from photoionized gas located at the surface of the obscuring torus \citep{G97,Roy98,G04}.
The size of $\sim 10$~mas corresponds to $r\sim 0.7$~pc, or $r/\rdust\sim 5-18$ for the
range of possible values of $L_{\rm bol}$ (see below). However, RPC photoionized gas is limited to 
a maximal value of $T_{\rm b}=2\times 10^5$~K at 5~GHz (Fig.~\ref{fig:Tb}). 

The solution for a uniform density photoionized gas is bound by an even lower value of $T_{\rm b}< 10^5$~K 
(Fig.~\ref{fig:const_n_Tb}).
We conclude that the observed high $T_{\rm b}$ value generally 
cannot be produced by free-free continuum from photoionized gas.
The $\sim 10$~mas core emission is therefore likely synchrotron emission. Given the low $T_{\rm b}$,
the synchrotron source is optically thin, and the intrinsic PL slope is therefore expected to
be significantly steeper than the observed $\alpha=-0.17$. The observed flat slope is potentially 
just an artefact produced by a free-free absorption observed below 4 GHz, 
as further discussed below.  

\begin{figure*}
	\includegraphics[width=2\columnwidth,trim = 4cm 0cm 4cm 1cm]{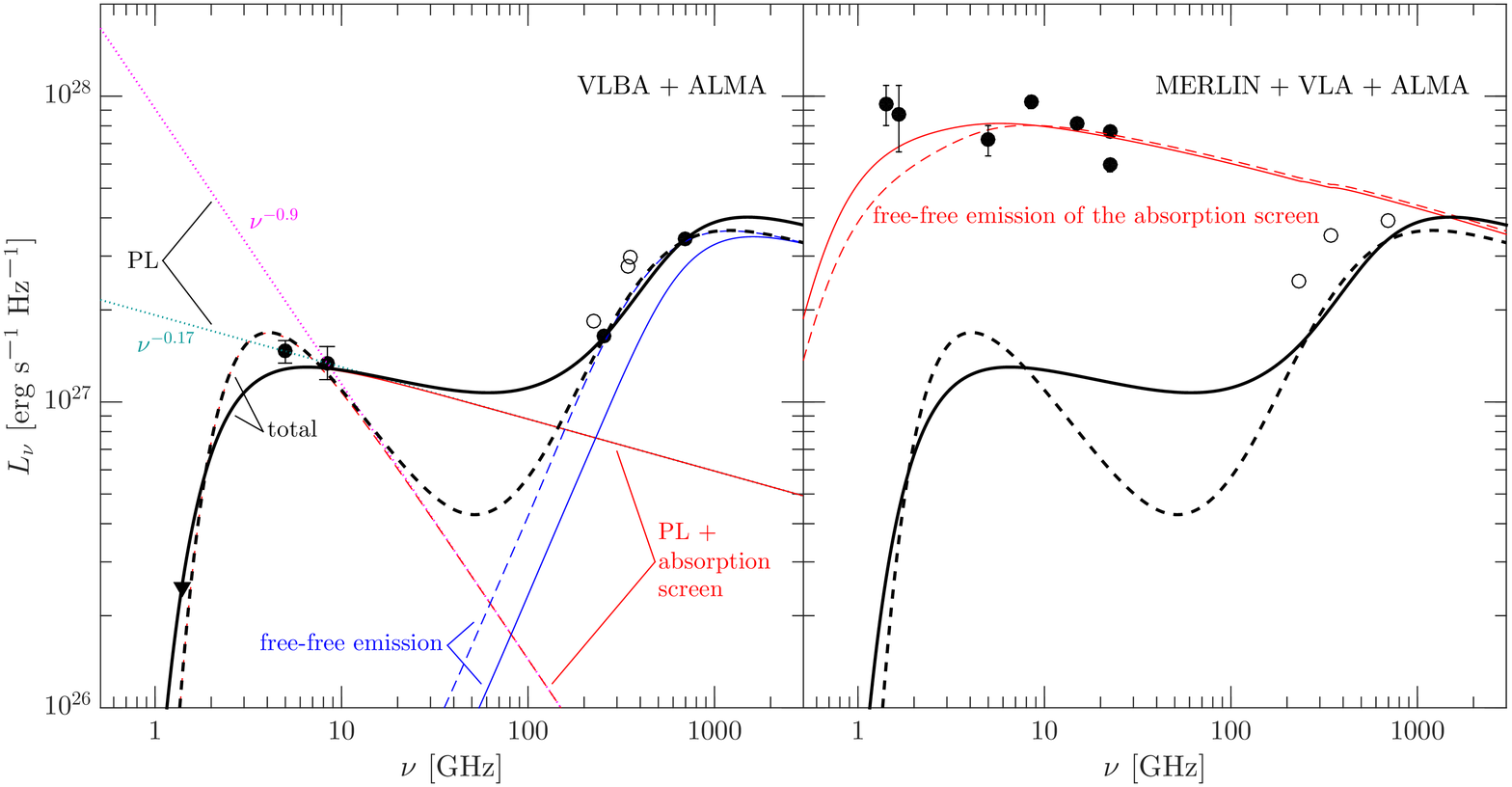}
	\caption{The mas scale (left panel) and sub-arcsec scale (right panel) 
		core radio emission in NGC 1068. {\bf Left panel:} VLBA observations below 10~GHz
		of the central emission feature (S1) which has a size of $\sim 5-15$~mas  \citep{G04}. 
		ALMA observations above 100~GHz, with a resolution of $\sim 20-60$~mas. The 
		filled dots are from \citet{Gar16} and \citet{Imp19}, and the circles from \citet{Inoue20}. 
		The sharp spectral turnover below 5~GHz indicates a free-free absorption screen. 
		The absorbed
		PL may be as flat as $-0.17$ (dotted cyan line) with a dusty RPC absorber located at
		$r/\rdust=975$ (solid black line), or it can be a steep PL with a slope of $-0.90$ 
		(dotted magenta line) with an absorber at $r/\rdust=915$ (dashed line). 
		The excess emission above 100~GHz is well fit by  
		free-free emission of RPC photoionized dustless gas, located either at $r/\rdust=3.1$
		with $\Omega=0.025$ for $\alpha=-0.17$ (solid lines), or at $r/\rdust=4.2$ with $\Omega=0.03$
		for $\alpha=-0.9$ (dashed lines). {\bf Right panel:} the more extended nuclear emission
		on scales of $100-200$~mas, derived from MERLIN and VLA observations \citep{G96}. 
		The relative strength of this more extended free-free emission 
		implies $\Omega\sim 0.1-1$, which is comparable to the
		values expected from the estimated distance and size of the free-free absorbing screen.}
	
	\label{fig:NGC1068}
\end{figure*}

Figure~\ref{fig:NGC1068}, left panel, shows the observed VLBA luminosity density at 5 and 8.4~GHz. 
There is a sharp spectral turnover below 5~GHz, with only an upper limit
at 1.4~GHz. The resolution matched flux densities at 1.4, 5 and 8.4~GHz are 
$<0.7, 5.9$ and 5.4~mJy, respectively \citep{G04}, implying slopes of $\alpha_{1.4-5}>1.67$ and $\alpha_{5-8.4}=-0.17$.
The observed spectral turnover from above to below 5~GHz is too sharp to match a self-absorbed free-free source 
(Fig.~\ref{fig:ill_side_nu}), and requires the presence of a foreground absorbing screen
(Fig.~\ref{fig:abs_only}). Assuming an unabsorbed PL source with the observed slope of $\alpha=-0.17$
extends down to 1.4~GHz, 
implies an unabsorbed 1.4~GHz flux of 7.3~mJy, and therefore $\tau_{1.4~{\rm GHz}}\ge 2.3$.
This optical depth is consistent
with a dusty RPC absorber located at $r/\rdust=975$. Since free-free absorption follows
$\tau_\nu\propto \nu^{-2}$ (eq.~\ref{eq:abs_co_approx}), this screen also produces $\tau_{5~{\rm GHz}}\ge 0.18$, and
$\tau_{8.4~{\rm GHz}}\ge 0.06$, so the observed value of $\alpha_{5-8.4}=-0.17$ is also affected by the free-free absorbing screen.
The unabsorbed $\alpha_{5-8.4}$ is steeper by $\Delta \alpha=0.24$ than the observed slope 
(see Fig.~\ref{fig:NGC1068}). Thus, the implied 
intrinsic PL slope is $\alpha=-0.41$, which is too steep to be free-free PL emission,
regardless of the gas heating mechanism.

The intrinsic slope of the absorbed PL source can also be significantly steeper, 
if the absorber is located slightly
inwards. For example, the observed $\alpha_{5-8.4}=-0.17$ can be produced 
by a PL source with $\alpha=-0.9$, a typical slope for an optically thin
synchrotron source,  if the absorber is located at $r/\rdust=915$ ($\Delta \alpha=0.73$), 
as demonstrated in Fig.~\ref{fig:NGC1068}.

Thus, the observed sharp spectral break below 5~GHz requires a foreground 
free-free absorbing screen. The absorption below 5~GHz implies that the 
absorption corrected $\alpha_{5-8.4}$ is steeper
than observed, which excludes free-free emission as the origin of the observed PL 
at $5-8.4$~GHz.  

The physical distance of the foreground screen, $r$, which resides at $r/\rdust=975$, 
can be derived from the value of $\rdust$, which is set by $L_{\rm bol}$. In NGC~1068,
$L_{\rm bol}=0.4-4.7\times 10^{45}$~erg~s$^{-1}$ \citep{Gravity20}, which implies
$\rdust=0.04-0.14$~pc (eq.~\ref{eq:rdust}). The absorber then resides at $r\sim 40-140$~pc.
If this is also the typical lateral dimension of the absorber, 
it corresponds to an angular scale of $0.5-2$~arcsec. This large extent is consistent with
the similar free-free absorption also observed in component C, which resides 0.3 arcsec
to the north of the centre \citep{G04}. 
The VLA observations of component C at
$8-22.5$~GHz reveal a steep PL with  $\alpha_{8-22.5}=-0.67$ \citep{G96}, 
but it shows $\alpha_{5-8.4}=-0.23$ \citep{G04}, again consistent with the spectral flattening 
close to the free-free cutoff observed below 5~GHz, which is similar to the observed VLBA scale absorption.

The free-free absorber screen should produces some 
free-free emission; is this emission detectable? 
RPC predicts a well defined value for the free-free $L_\nu/L_{\rm bol}$  
at a given $r/\rdust$ and $\Omega$ (Fig.~\ref{fig:ill_side_nu}). 
For the above values of $L_{\rm bol}$, $r/\rdust$ and $\Omega$ ($=0.3$), the RPC model predicts
$L_{\rm 5~GHz}=2\times 10^{27}-2.4\times 10^{28}$~erg~s$^{-1}$~Hz$^{-1}$. The
observed VLBA value is $L_{\rm 5~GHz}=1.5\times 10^{27}$~erg~s$^{-1}$~Hz$^{-1}$.
However, the observed VLBA scale emission is not relevant, as the absorber screen emission is expected to extend
over an angular scale of $\sim 0.5$ to 2 arcsec, and the screen emission within the 
tiny VLBA 
scale of $\sim 10$~mas is only $\sim 10^{24}$~erg~s$^{-1}$~Hz$^{-1}$. Thus, the absorber free-free
emission on the VLBA scale is negligible compared
to the compact source emission. Is there evidence for free-free emission on larger scales?

Figure~\ref{fig:NGC1068}, right panel, shows the observed emission on scales of $\sim 100-200$~mas 
from the nucleus. The spectral slope is $\alpha=-0.08$ \citep{G96}, which matches well
free-free emission. The plot also shows the free-free emission expected from 
the above derived position of the free-free absorber,
$r/\rdust=915$ and 975. The amplitude of the free-free luminosity implies 
that the luminosity absorbed by the screen is $ \Omega L_{\rm bol}=4.9\times 10^{44}$~erg~s$^{-1}$.
The estimated $L_{\rm bol}=0.4-4.7\times 10^{45}$~erg~s$^{-1}$ 
implies $\Omega\sim 0.1-1$ for the screen. The value of $\Omega$
can also be estimated from the known geometry of the system.
The observed projected size of the free-free emitter is $h\sim 10$~pc, 
and its distance from the centre, based on RPC is $r\sim 40-140$~pc (see above).
If the absorber forms a torus like structure then  $\Omega\sim h/r$, or $\Omega\sim 0.07-0.25$, which  overlaps $\Omega\sim 0.1-1$ derived above from the free-free emission amplitude.

To summarize, the $\sim 10$~mas flat PL nuclear source is too bright and too steep 
to be free-free emission, and is likely optically thin synchrotron emission. 
The emission is clearly absorbed by a free-free
screen located at $r/\rdust\sim 1000$. The free-free emission of this screen is apparently 
detected on larger scales of $100-200$~mas. This large scale RPC free-free emitting screen is 
likely also the source of the spatially resolved X-ray emission lines in NGC~1068 \citep{Ogle03}.
The X-ray emission lines luminosity and their relative strengths are also well matched by RPC emission \citep{paperV}.

\subsection{Exclusion of free-free emission from hot gas}

Above we excluded photoionization, either of RPC gas or of uniform density gas, 
as the heating mechanism for the $\sim 10$~mas PL component observed 
in NGC 1068. Can radio emission with $T_{\rm b}\sim 4\times 10^6$~K generally be produced 
by free-free emission of hot gas? Below we show that in AGN such hot gas
over predicts the observed X-ray emission.

In order to show that, we first estimate the required gas $T_{\rm e}$, in order to produce the 
observed $T_{\rm b}$. If the gas is optically thick at 5~GHz, then $T_{\rm e}=T_{\rm b}$. But, the thermal radio emission of optically thick gas is blackbody, 
that is a PL with $\alpha=2$, while the observed slope is $\alpha_{5-8.4}=-0.17\pm 0.24$ \citep{G04}. 
The hot gas therefore needs to be optically thin, in which case $T_{\rm e}=T_{\rm b}/\tau_\nu$.
The minimal $T_{\rm e}$ requires the largest possible $\tau_\nu$,
which is consistent with the observed $\alpha_{5-8.4}$.
The largest acceptable value is $\tau_{\rm 5~GHz}=0.1$, as it produces free-free emission with $\alpha_{5-8.4}=0.06$, which is within the acceptable range of $\alpha_{5-8.4}=-0.17\pm 0.24$. 
Thus, the observed $T_{\rm b}\sim 4\times 10^6$~K and the constraint $\tau_\nu<0.1$
implies $T_{\rm e}> 4\times 10^7$~K. 

\begin{figure}
	\includegraphics[width=\columnwidth]{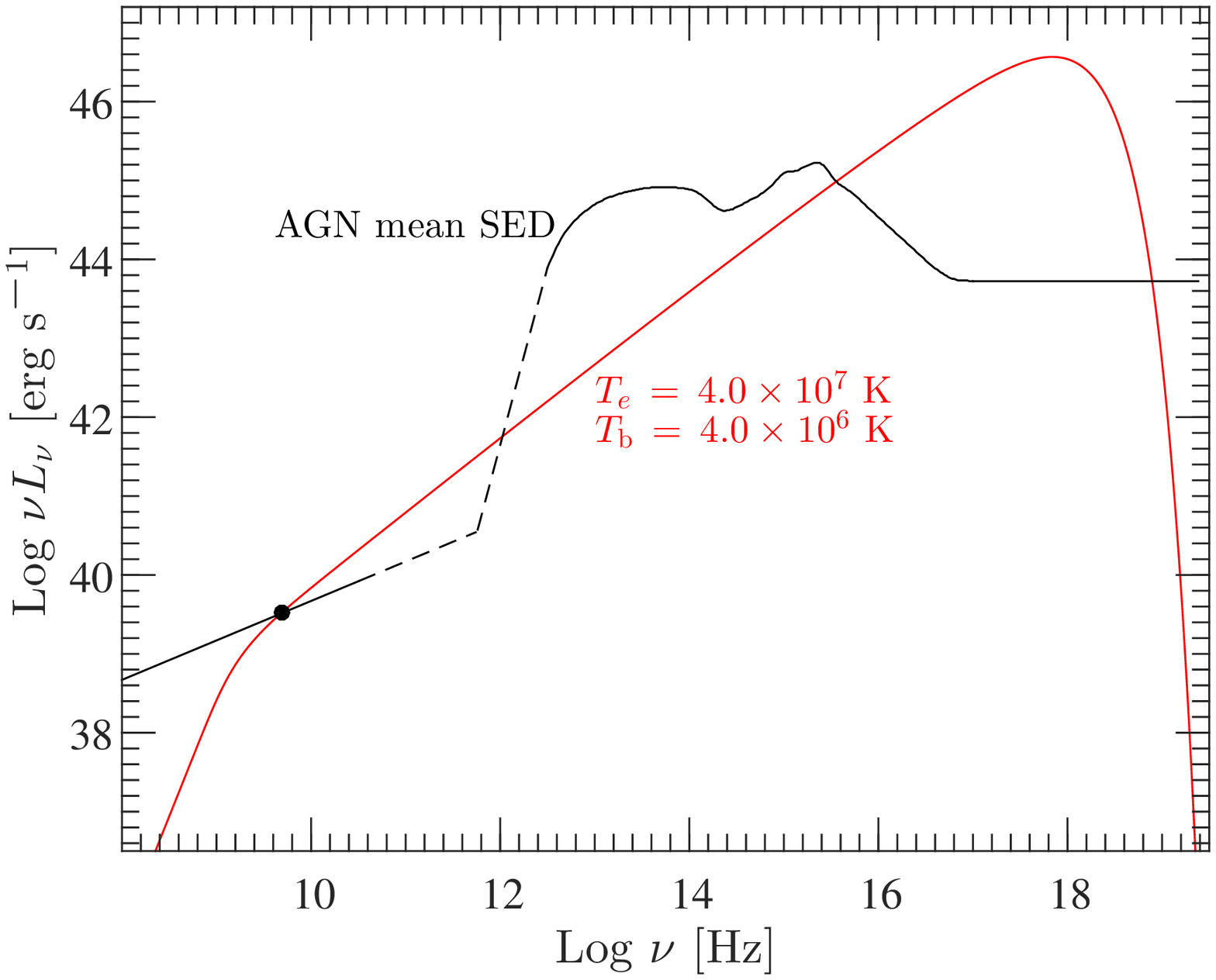}
	\caption{The free-free emission of hot gas 
		with $T_{\rm e}=4\times 10^7$~K (red line), which produces a high-brightness-temperature radio emission
		with $T_{\rm b}=4\times 10^6$~K at 5~GHz, as observed in NGC~1068. Such a source produces 
		$L_{\rm 2~keV}/\nu L_{\rm 5~GHz}=10^7$, which is a factor of 600 larger than typically observed
		in radio quiet AGN \citep{LB08}. Thus, if a significant fraction of the observed radio emission at 5~GHz is 
		characterised by $T_{\rm b}> 10^6$~K, the emission  
		cannot be thermal free-free emission of hot gas, and is inevitably synchrotron emission. 
	}
	\label{fig:NGC1068_SED}
\end{figure}

Can the core radio emission in NGC~1068, and generally the radio emission 
in radio-quiet AGN, be produced by free-free emission from $T_{\rm e}> 4\times 10^7$~K gas?
Figure~\ref{fig:NGC1068_SED} presents the calculated free-free emission of gas at $T_{\rm e} =4\times 10^7$~K,
with $\tau_{\rm 5~GHz}=0.1$, and thus $T_{\rm b}({\rm 5~GHz})=4\times 10^6$~K, as observed in the core of
NGC~1068. 
The free-free emission is overlaid 
on the mean SED of type 1 AGN \citep{richards_etal06}, by matching their $\nu L_{\rm 5~GHz}$. The free-free emission peaks at $\nu \sim 10^{18}$~Hz, 
or an energy of $h\nu\sim 3$~keV.
Clearly, the implied free-free X-ray luminosity 
far exceeds the observed X-ray luminosity. Specifically, the free-free
emission gives $L_{\rm 2~keV}/\nu L_{\rm 5~GHz}\sim 10^7$, compared to the observed mean AGN SED 
which is characterised by $L_{\rm X-ray}/\nu L_{\rm Radio}\sim 10^5$, or equivalently 
$L_{\rm 2~keV}/\nu L_{\rm 5~GHz}\sim 1.6\times 10^4$ (using the X-ray 
bolometric correction factor $L_{\rm X-ray}=6.25L_{\rm 2~keV}$ from \citealt{LB08}).
The hot free-free emitting gas over produces the X-ray luminosity by a factor of $\sim 600$.

The above argument applies for an average unobscured AGN. In NGC~1068 the X-ray emission is obscured, 
and the absorption corrected value is estimated in the range of
$L_{\rm 2~keV}=0.2-4\times 10^{43}$~erg~s$^{-1}$, assuming a PL with a slope of $-1$
(\citealt{Gravity20}, and references therein).
The luminosity of the $T_{\rm b}=4\times 10^6$~K source is $\nu L_{\rm 5~GHz}=7.5\times 10^{36}$~erg~s$^{-1}$, 
and if this is free-free emission of hot gas, it implies 
$L_{\rm 2~keV}=7.5\times 10^{43}$~erg~s$^{-1}$, which is about twice 
the estimated intrinsic $L_{\rm 2~keV}$. 
 
An additional difficulty to accommodate the free-free X-ray emission is its spectral shape. 
The observed X-ray spectral shape in AGN is a PL with a slope of $\sim -1$,
in contrast with the free-free SED of a PL with a slope of $\sim -0.1$ with a thermal cutoff at
$\sim 3$~keV. In addition, the values of $\tau_{\rm 5~GHz}$ and $L_{\rm 5~GHz}$ in NGC~1068,
together with eqs.~\ref{eq:L_ratio} and \ref{eq:abs_co_approx} allow to derive the density and size of the free-free
emitting region, which gives $n_{\rm e}=2\times 10^6$~cm$^{-3}$, and a size of 0.6~pc. 
The size is in sharp contrast with the typical size of the X-ray emitting
region in AGN of $\sim 10^{-3}$~pc. 

To summarize, radio emission component in AGN with $T_{\rm b}>10^6$~K at 5~GHz 
can be free-free emission only if the radio is weak enough to satisfy 
$\nu L_{\rm 5~GHz}/L_{\rm 2~keV}< 10^{-7}$, so it does not dominate
the X-ray emission. Such a source produces less than $\sim 1$ per cent of the radio emission
at 5~GHz. 
Otherwise, the observed X-ray emission becomes
dominated by the free-free emission, which is inconsistent with the observed spectral slope
 and size of the X-ray emitting region (likely produced in a corona above the innermost accretion
disc). Significant radio emission with $T_{\rm b}>10^6$~K in radio-quiet AGN is therefore likely 
to be synchrotron emission.

A synchrotron source requires relativistic electrons with $\gamma\gg 1$, and
thus kinetic energy $\gg m_ec^2$. If the source is 
optically thick it will inevitably have $T_{\rm b}\gg 10^9$~K. 
A synchrotron source with $T_{\rm b}<10^9$~K
is most likely optically thin, and is thus expected to produce steep PL emission. 
In NGC~1068, the observed flat $\alpha_{5-8.4}$ is most likely just an artefact of the 
free-free absorption screen, which becomes dominant around 5~GHz. 
The free-free absorption corrected radio slope is $\alpha_{5-8.4}<-0.5$, 
which is consistent with optically thin synchrotron emission.

\subsection{Free-free emission in other active galaxies}\label{sec:other}

\cite{Mundell00} searched for free-free continuum on the mas scale in five nearby Seyfert galaxies.
In four of these, $T_{\rm b}>10^8$~K, which excludes free-free emission (see above). In NGC~4388, the non detection
of mas scale emission places an upper limit of $T_{\rm b}<2.2\times 10^6$~K, which is consistent with 
free-free emission. However, followup VLBI observations at 1.6 and 5~GHz 
by \cite{GP09} find a spectral slope of $\alpha<-0.7$, which excludes core free-free emission.
The flat spectrum emission detected on $\sim 50-100$~mas scale in NGC~4388 with MERLIN,
may well be produced by free-free emission on scales of $5-1$~pc \citep{Mundell00}.

Of the four additional faint Seyfert nuclei observed by \cite{GP09}, NGC~5033 shows the expected 
free-free slope of $\alpha=-0.1\pm 0.1$, but its $T_{\rm b}>1.3\times 10^7$~K excludes the free-free
interpretation. 

The most, and maybe only, convincing detection of free-free emission is from the nucleus  
of Cygnus~A. \cite{C19} detected flat PL emission, with $\alpha=-0.1$ from 18 to 48~GHz, 
from a torus structure with a size of $\sim$500~pc. The torus resides 
perpendicular to the radio loud jet. The observed radio intensity at 34~GHz gives 
$T_{\rm b}=240$~K at 100~mas, which corresponds to a distance of 120~pc from the centre. 
The bolometric luminosity of Cygnus~A is $0.5-2\times10^{46}$~erg~s$^{-1}$
\citep{Tad03}, which implies $\rdust\sim 0.2$~pc, and thus 
$T_{\rm b}$ is measured at $r/\rdust \sim 600$.
An RPC free-free emitting dusty gas at $r/\rdust \sim 600$ is expected to remain
optically thin down to $\sim 1.4$~GHz (eq.~\ref{eq:v_thick_RPC}), which is consistent with
the observed optically thin emission down to 18~GHz \citep{C19}. 
However, since 
$\tau_\nu\propto \nu^{-2}$, we expect that $\tau_{34~{\rm GHz}}\sim 2\times 10^{-3}$,
which implies a predicted $T_{\rm b}=\tau T_{\rm e}\sim 20$~K for photoionized gas 
(see also Fig.~\ref{fig:Tb}),
in contrast with the observed value of 240~K. Also, for an estimated total flux
density from the torus of $\sim$2~mJy, we get that $L_\nu/L_{\rm bol}=5\times 10^{-17}$
at 34~GHz, compared to the predicted RPC value of $L_\nu/L_{\rm bol}=4\times 10^{-18}$
(Fig.~\ref{fig:ill_side_nu}). Again, the predicted free-free emission from AGN photoionized
gas is a factor of $\sim 10$ too low. If the free-free emitting gas is dustless,
then $L_\nu/L_{\rm bol}$ is a factor $\sim 10$ larger (Fig.~\ref{fig:ill_side_nu}),
and is consistent with the observed ratio. Dustless gas at such a high 
$r/\rdust$ of $\sim 600$ would suggest the gas originates from a gas outflow 
which starts at $r/\rdust<1$, possibly related to the jet activity in this RL AGN.
Alternatively, other local gas heating mechanism, such as 
star formation, or mechanical jet heating, may also boost the free-free intensity to the observed value.

The two critical points for the successful detection of free-free emission in Cygnus~A
are: first, the high frequencies used; and second, the high sensitivity of the observations. The high frequencies 
enhance the probability to detect the flat spectrum free-free emission, against extended
steep non thermal emission.
The high frequencies also provide a higher angular resolution, which reduces the
contamination from extended sources on larger scale. The high sensitivity is required
given the expected low $T_{\rm b}$ of free-free emission. Specifically, in Cygnus~A the possible free-free emission is
at the 1~mJy level, compared to the total radio emission of
$\sim 100$~Jy in this luminous nearby radio loud AGN.

\subsection{Free-free emission from the Broad Line Region}\label{sec:BLR}

The free-free emission likely detected in NGC~1068 (Section~\ref{sec:ngc1068}) 
 and Cygnus~A (Section~\ref{sec:other}) is produced at
$r/\rdust \sim 600-1000$, which corresponds to the NLR scale.
Free-free emission is also predicted to be produced by the photoionized gas in
the BLR, but since free-free self-absorption is more prominent at the BLR,  
its detection requires observations above 100~GHz (see Fig.~\ref{fig:ill_side_nu}).

A unique signature of the free-free emission from photoionized gas in AGN 
is the expected spectral break between the
NLR free-free emission and the BLR free-free emission. The break occurs due to the dust sublimation at
$r/\rdust$ of a few, and the consequent sharp rise in $L_\nu$ by a factor
of $\sim 10$ around 100~GHz (Fig.~\ref{fig:ill_side_nu}).
The detection of mm emission in RQ AGN is becoming feasible now with the ALMA array, which
may already provide detections of the expected mm spectral break and the BLR free-free emission. 

Figure~\ref{fig:NGC1068} presents high frequency ALMA observations of NGC 1068 at 256~GHz 
\citep{Imp19}, and 694~GHz \citep{Gar16}, which show a rising emission component
with corresponding fluxes of 6.6~mJy and 13.8~mJy, implying a local spectral slope 
of $\alpha=0.74$.
This component resides significantly above an extrapolation of the VLBA low frequency PL component, 
which contributes at most about half the ALMA detected flux, when extrapolated to 256~GHz. 
The ALMA beam size used at 256~GHz is $\sim 20$~mas,  and $\sim 60$~mas at 694~GHz \citep{Gar16,Imp19}, 
which raises the possibility that the flux rise is a beam size effect for a diffuse source. 
However, the fluxes measured by 
\cite{Inoue20} above 100~GHz, with a beam size of $\sim 30$~mas, lead to a similar spectral shape, which suggests 
the flux rise is real and not a beam size effect.

The steeply rising excess emission above 100~GHz is remarkably similar 
to the predicted free-free emission of RPC gas on the BLR scale, which is characterised by
$\nu_{\rm thick}>100$~GHz (Fig.~\ref{fig:ill_side_nu}).
The excess flux is well described by dustless RPC gas with $\Omega=0.025$ located at $r/\rdust=3.1$, assuming an underlying
flat PL $\alpha=-0.17$ component, or $\Omega=0.03$ and $r/\rdust=4.2$ for an assumed steep PL component with
$\alpha=-0.9$. For dusty gas free-free emission, the corresponding best fit parameters are
$\Omega=0.23$ and $r/\rdust=1.1$ for $\alpha=-0.17$, and 
$\Omega=0.28$ and $r/\rdust=1.7$ for $\alpha=-0.9$. The smaller value of $r$ 
for dusty gas is due to the lower $\tau$ of dusty RPC gas at a given $r/\rdust$,
and the larger $\Omega$ is due to the weaker emission of dusty gas (Fig.~\ref{fig:ill_side_nu}). 
The dustless solution is favoured by the dust physics and
by the GRAVITY observation of a compact dusty disc at the same location \citep{Gravity20}, 
as further discussed below.

Excess mm emission, above the cm PL extrapolation, 
is commonly observed in RQ AGN \citep{Doi05, Doi11, behar_etal15,behar_etal18}, and sometimes 
shows an inverted slope of a rising flux with frequency. For example, \cite{ID18} find
a flux rise above 20~GHz in  NGC~985, and above 50~GHz in IC~4329A. However,
ALMA observations above 100~GHz 
show a spectral steepening to $-0.37$ in NGC~985, and to $-0.51$ in IC~4329A
\citep[tables 1 and 2]{ID18}, which clearly excludes free-free emission,
and suggests the mm excess emission is produced by compact optically thick synchrotron emission.

\subsection{\LOIII\ versus $L_{\rm radio}$, predicted versus observed}

The RPC effect of photoionized gas leads to a unique ionization and density structure of the gas at the NLR and the BLR \citep{paperI,paperII}, and thus to a unique solution for the line and continuum emission
for a given incident flux.
One can therefore use the line luminosity to predict the expected free-free luminosity,
and compare it to the observed relation. The free-free emission from the BLR is restricted
to $\nu > 100$~GHz, which is not readily available yet for RQ AGN. At $\nu < 100$~GHz,
the free-free emission originates from the NLR, and we therefore use below
the narrow line luminosity, specifically the \OIII~$\lambda5007$ line, which is the strongest line from the NLR, to explore the free-free versus line luminosity relation.

\begin{figure*}		
\includegraphics[width=2\columnwidth]{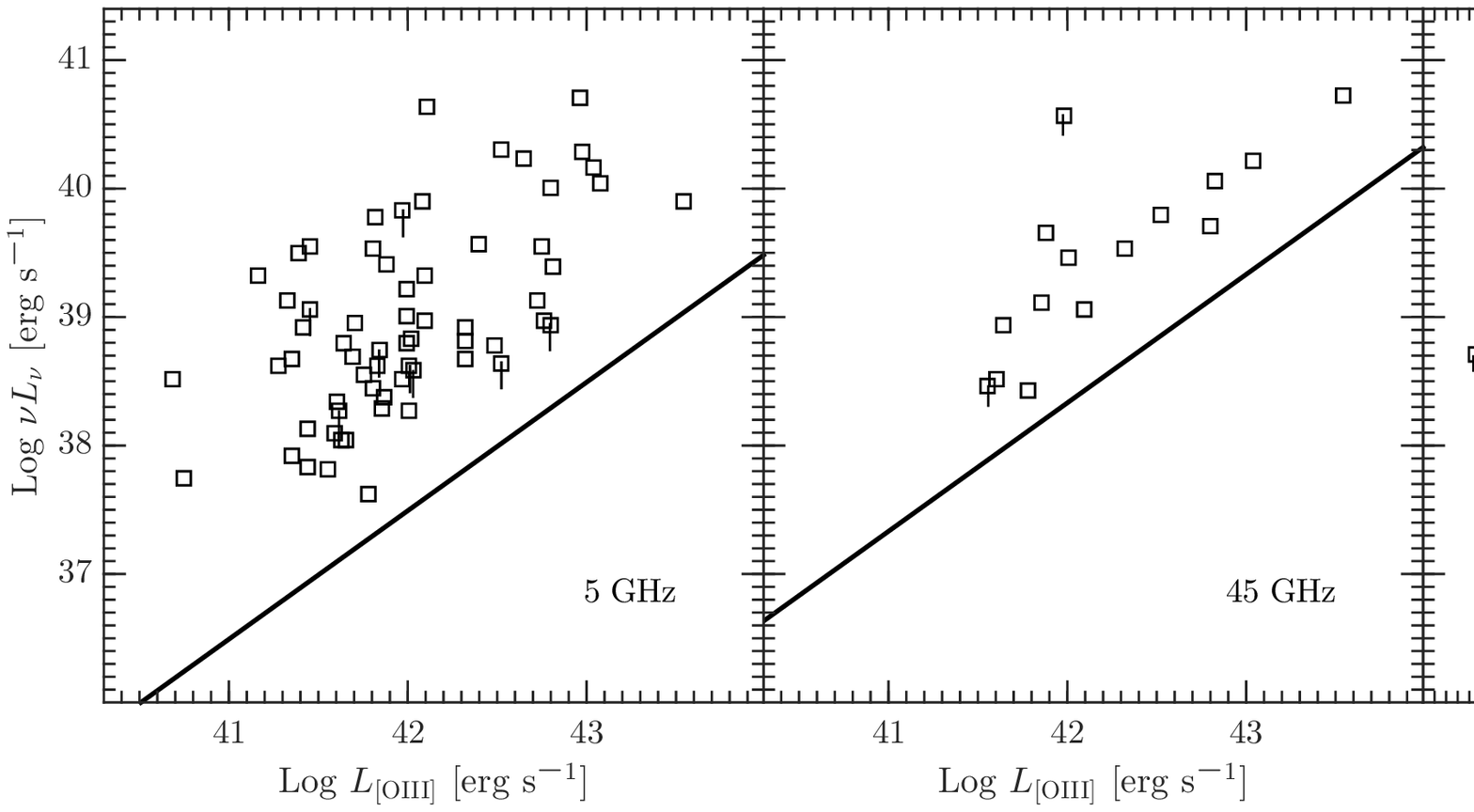}
\caption{ 
The predicted $\nu L_\nu$ versus \LOIII\ from Cloudy model results, compared to the observed distributions at $\nu=5$, 45 and $95-100$~GHz.
{\bf Left panel: } 
The  observed \Lfive\ versus \LOIII\ distribution for the radio quiet PG quasars
(open squares), compared to the RPC based thermal free-free radio emission (solid line).  The radio emission
of all objects resides well above the predicted free-free \Lfive, suggesting the radio is dominated
by synchrotron emission, as also indicated by the observed range of spectral slopes. The objects
with the lowest \Lfive, at a given \LOIII, are typically only a factor of $\sim 5$ above the predicted free-free emission. 
{\bf Middle panel: }
Same as the left panel for \Lfourtyfive . 
 The observed \Lfourtyfive\ of all objects is above the predicted free-free emission, but the median excess flux is now only a factor of $\sim 5$,
compared to a median excess of $\sim 20$ at 5~GHz (left panel).
{\bf Right panel: }
Same as the left panel for  $L_{95, 100{\rm ~GHz}}$. The objects are hard X-ray selected  mostly nearby AGN, which are either obscured or unobscured. The observed distribution resides slightly closer to the predicted free-free emission. Note that some of these hard X-ray selected objects may be displaced due to \OIII\ absorption.
}

\label{fig:R_vs_OIII}	        
\end{figure*}

Figure~\ref{fig:R_vs_OIII} compares the predicted and observed relation between the \OIII~$\lambda5007$ emission line luminosity and $L_\nu$ at $\nu=5, 45$ and $95-100$~GHz, in radio quiet 
AGN. The left panel presents \Lfive\ versus \LOIII\ for the RQ PG quasars. 
The radio luminosities are taken from \cite{K89}, and the \OIII~$\lambda5007$ emission line luminosities are
taken from \cite{BG92}.\footnote{The luminosities are derived using: $H_0 = 67.7$~\kms~Mpc$^{-1}$, $\Omega_{\rm m} =0.31$ and $\Omega_\Lambda=0.69$.} The predicted relation is derived from the RPC dusty gas calculations  for $r/\rdust = 10^{3.5}-10^5$ (e.g.\ $r=0.6-20$~kpc for $L_{46}=1$), the range of distances where 
most of the \OIII~$\lambda5007$ line emission is expected to be produced (\citealt{paperI}, fig.~6).  We use the value of \Lfive\ emitted from the illuminated side of the slab; and the total \LOIII, as calculated by Cloudy, which mostly comes from the illuminated side. The Cloudy results imply a relation of
	\begin{equation}
		\log \nu\Lfive = \log \LOIII -4.51\,. \label{eq:predcit_L5G} 
	\end{equation}
The observed \Lfive\ versus \LOIII\ distribution of the RQ PG quasars lies well above the predicted relation, with a median 
observed excess of \Lfive\ above the free-free value of about a factor of 20. The objects
with the lowest \Lfive, at a given \LOIII, are still a factor of $\sim 5$ above the free-free relation.
Indeed, the observed radio spectral slope is typically either significantly steeper
or flatter than $\alpha=-0.1$ of optically thin free-free emission, suggesting  
synchrotron emission, which is either optically thin or optically thick \citep{laor_etal19}.

However, the predicted \Lfive\ is not always hopelessly swamped by the non thermal radio emission, in particular in those
objects at the
 lowest \Lfive\ for a given \LOIII. Also, since the free-free spectral slope is flat, 
it is expected to become more dominant at higher frequencies, in particular in objects dominated by a steep synchrotron emission.

What is the observed relation with \LOIII\ at higher radio frequencies?  
Figure~\ref{fig:R_vs_OIII}, middle panel, 
compares the observed and the predicted relation between \LOIII\  and \Lfourtyfive, for a small
subset of 15 RQ AGN, selected randomly from the PG quasar sample \citep{Baldi21}.  
The predicted relation based on RPC is 
	\begin{equation}
	\log \nu\Lfourtyfive = \log \LOIII -3.67\,. \label{eq:predcit_L45G}
	\end{equation}
The observed distribution of \Lfourtyfive\ versus \LOIII\ lies above the predicted relation, as found
above for \Lfive\  (Fig.~\ref{fig:R_vs_OIII}, left panel). However, the median excess flux is only a factor of $\sim 5$
above the predicted free-free emission, in contrast with the median excess flux of $\sim 20$ found above 
at 5~GHz.  
		
What is the observed relation at yet higher radio frequencies?  
Figure~\ref{fig:R_vs_OIII}, right panel, presents the observed relation between the radio luminosity at $\nu\simeq100$~GHz and \LOIII, for a hard X-ray selected sample of mostly lower luminosity nearby AGN, some of which are obscured. 
We adopt the measured flux at $\nu=95$~GHz from \citet{behar_etal15}, and at $\nu=100$~GHz from \citet{behar_etal18}. The corresponding values of \LOIII\ are from \citet{koss_etal17}.\footnote{The value of \LOIII\ for the object Ark~564 is adopted from \citet{Crenshaw_etal02}, since it is not covered by \citet{koss_etal17} (the value of the luminosity distance is adopted from \citealt{behar_etal15}). NGC~5106 is excluded from the sample since it is radio loud.} We also adopt from \citet{koss_etal17} the distance and $z$ of the sample objects, and convert the measured radio fluxes to $L_\nu$. The PL fit of the modelled free-free emission at $\nu=100$~GHz yields
	\begin{equation}
	\log \nu\Lhundred = \log \LOIII -3.36\, . \label{eq:predcit_L100G}
	\end{equation}
The observed distribution again reside above the free-free prediction, as  
found at 5 and 45~GHz. 
Objects at $\log \LOIII>41$ reside within a factor of $3-10$ above the predicted free-free emission,
while objects at $\log \LOIII<41$ have a significantly larger excess. This could be just a selection effect, as only the higher \Lhundred\ AGN are detectable when the sample extends to the lowest \LOIII. Since some of the objects are obscured AGN, their \LOIII\ may be offset to lower values, which will also increase the deviation from the free-free relation.

\subsection{The free-free continuum versus the observed SED}

\begin{figure}
		\includegraphics[width=\columnwidth]{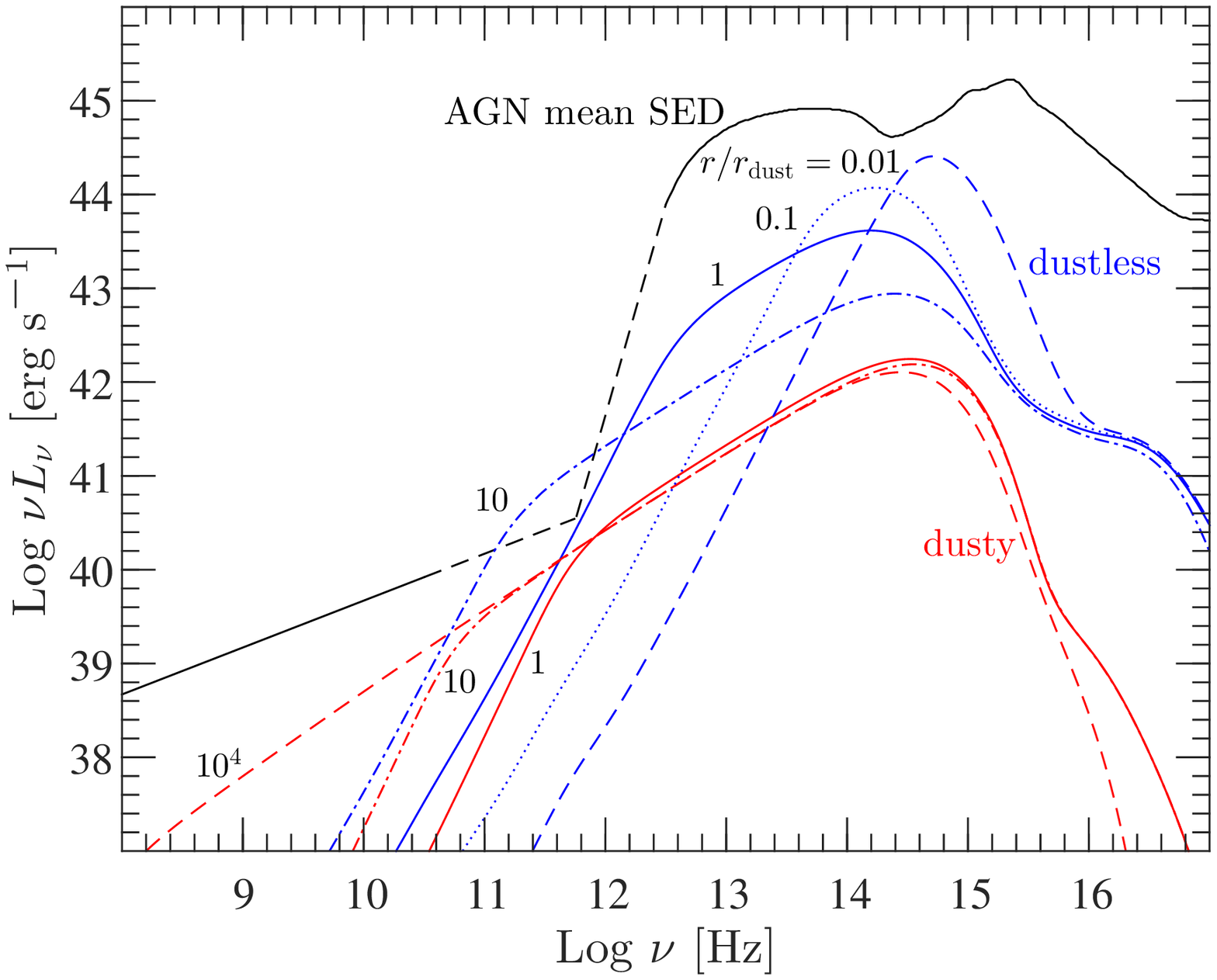}
		\caption{The RPC gas free-free continuum from gas located at 
various $r/\rdust$, 
compared to the mean observed AGN SED (black), scaled for $L_{\rm bol}=10^{46}$~\ergs. 
The RPC calculations are presented for dusty (red) and 
dustless (blue) gas. The free-free continuum of dustless  gas
which resides just outside the BLR ($r/\rdust\sim 10$) 
may be detectable at $\nu\approx 100 - 800$~GHz. Free-free emission at 1000~GHz ($300~\mu$m) may also be detectable,
despite the dominant cold dust emission, as the dust emission is extended on arcsec (kpc) scale, compared to the micro arcsec (sub pc BLR) scale free-free emission.}
		\label{fig:obs_SED}
	\end{figure}

Figure~\ref{fig:obs_SED} compares the predicted free-free continuum from RPC gas at 
various distances, with the observed mean SED of RQ AGN, scaled for 
$L_{\rm bol}=10^{46}$~\ergs. The mean SED is calculated as follows. In the range of 
$\nu=10^{12.5}-10^{17}$~Hz, we use the mean SED from \citet{richards_etal06}. 
We extend the SED to $\nu>10^{17}$~Hz assuming a constant $\nu L_{\nu}$, and to
the sub-mm
regime using the relation $L_{\nu}\propto\nu^{3.5}$ \citep{S16}.
For the radio range, we adopt a PL  $L_{\nu}\propto\nu^{-0.5}$, 
and use the normalisation $\nu L_{\nu}(5{\rm~GHz})/\nu L_{\nu}(10^{17}{\rm~Hz})=6.25\times 10^{-5}$ \citep{LB08}, which corresponds to a radio loudness parameter of $R=0.6$.
For the RPC models, we present the free-free emitted from the illuminated side for $\Omega=0.3$. 

The predicted free-free continuum of 
most models lie well below the observed SED at most frequencies. The free-free SED is potentially significant, or even dominant, only at $\nu\approx 100-300$~GHz, located at the presumed sharp spectral break between
the extrapolation of the radio cm band emission to the mm regime, and the extrapolation of the cold dust FIR 
emission to the sub mm regime.
The free-free continuum of dusty gas models becomes comparable to the estimated mean SED luminosity at $\sim 500$~GHz. 
The free-free emission of dustless gas is a factor of $\sim$10 stronger than dusty gas,
and may dominate the assumed SED above 100~GHz, if significant dustless gas can be found outside the BLR
at $r/\rdust \sim 10$. Free-free emission from dustless gas inside the BLR ($r/\rdust < 1$) can also be detected, 
despite the dominance of the sub mm tail of the cold dust emission at $\nu >500$~GHz, as the free-free 
source is far more compact ($\ll $1~mas) than the cold dust ($\sim 1$~arcsec).  The difference in $T_{\rm b}$ is
also dramatic. In both dusty and dustless RPC gas, we expect free-free emission with 
$T_{\rm b}\sim 2-3\times 10^4$~K (Fig.~\ref{fig:Tb}). The dust emission at $\nu >500$~GHz 
comes from the coldest dust, where $T_{\rm e}<100$~K. Also, the dust optical depth is inevitably very low.
For example, at 600~GHz we expect $\tau \sim 0.07$ even for a very high mean column of $10^{24}$~cm$^{-2}$
\citep[fig.~6]{laor_draine93}.
Since $T_{\rm b}=\tau T_{\rm e}$ for $\tau<1$, the $T_{\rm b}$ of dust emission will be smaller by a factor $>10^3$ than the $T_{\rm b}$
of the BLR-scale free-free emission.

Fig.~\ref{fig:obs_SED} also explores the predicted free-free emission of photoionized gas which
extends down to $r/\rdust=0.01$, well inwards of the standard BLR. As $r/\rdust$ decreases the gas becomes optically thick at higher $\nu$ (Fig.~\ref{fig:abs_only}). The peak free-free $L_\nu$ increases 
with decreasing $r/\rdust$
(see also Fig.~\ref{fig:ill_side_nu}). The emission at $\nu\gtrsim 10^{16}$~Hz is roughly independent of $r/\rdust$ 
(Fig.~\ref{fig:obs_SED}). This $\nu$-range roughly corresponds to $T\approx h\nu/k\gtrsim 5\times10^5$~K,
and corresponds to the free-free emission from the hottest, optically thin, surface layer where the RPC
gas reaches the Compton temperature at all distances \citep{paperI, paperII}. The emission from this
optically thin layer is $L_{\nu}\propto n^2 r^2 l$, which  is roughly independent of $r/\rdust$ for RPC gas. 
Gas located at $r/\rdust=0.01$ reaches a maximal density of $n\sim 10^{15}$~\cmt\ in an RPC slab \citep{paperII},
which is the high-density limit of the photoionization code Cloudy. 
Thus, we refrain from extending the calculation to a smaller value of $r/\rdust$. Since the typical
$\rdust \sim 10^4 r_{\rm g}$, where $r_{\rm g}=GM_{\rm BH}/c^2$ is the gravitational radius, a value of
$r/\rdust<0.01$ corresponds to $r<100r_{\rm g}$ which corresponds to the inner parts of the accretion disc, and the
assumption of a point source illumination breaks down.

\section{Discussion} \label{sec:discussion}

\subsection{Absorption}

Radio absorption is a powerful tool for the detection of both neutral and ionized gas is AGN. 
Neutral gas can be probed through the \HI~21~cm absorption line \citep[e.g.][]{M04}, while ionized gas
can be discovered through its free-free absorption (Sections~\ref{sec:ff_abs} and 
\ref{sec:obs_ff_abs}). The great advantage of the radio is that it allows to spatially 
resolve the structure of the absorbing medium on mas, i.e.\ pc, scale.
Such a spatial resolution of thermal emission and absorption  
is generally not achievable in other bands (with the exclusion 
of the recent outstanding GRAVITY results, e.g.\ \citealt{Gravity20}).
 The significant limitation of the radio method is that it requires a very bright
radio background source on mas scale. So, this tool is restricted to the small subset of AGN which
are both radio loud, and have a high $T_{\rm b}$ compact source. In addition, the spatial distribution 
of the absorbing medium is limited to gas which happened to resides in front of the radio source \citep[e.g.][]{V94, U99a, G99, W00, Jones01, Kameno03}.

The free-free absorbing gas on pc scale can be heated and ionized through various mechanisms. It can be viscous
heating, if it resides within the accretion disc. It can be shock heating due to a wind
or jet interaction with the ambient medium. It can be photoionized by the AGN,
which is inevitable if the gas is exposed to the central source. In this case, the RPC effect is also inevitable, but it is not necessarily the dominant gas compression mechanism, as the gas may be 
magnetic pressure confined, or thermal pressure confined by a hot and dilute medium.

The unique advantage of the RPC solution is that it provides a unique prediction for 
$\tau_\nu$ as a function of the distance 
of the absorbing gas from the ionizing source (Section~\ref{sec:ff_abs}).
This is in contrast with 
non RPC photoionization solutions \citep[e.g.][]{Ulvestad81, KL89}, where $\tau_\nu$
depends on the unknown gas density. Thus, an observational determination
of $\tau_\nu$ does not yield the absorber distance. In contrast, the
 RPC effects produces such a relation. Specifically, absorbers on a scale of $r/\rdust\sim 1000$, 
or 10 to 100~pc for AGN at $\Lbol\sim 10^{44}$ to  $10^{46}$~\ergs, 
has $\nu_{\rm thick}\sim 1$~GHz, and on scales of $1-10$~pc will show
$\nu_{\rm thick}\sim 10$~GHz. 

In the well studied case of NGC~1275 \citep{J96, U99, P99, V03}, we find that indeed
the observed spatially resolved $\nu_{\rm thick}$ matches the RPC prediction,  
based on the luminosity of NGC~1275 and the spatially resolved physical 
distance of the absorber. Similar agreements are found for the spatially resolved free-free absorption
in NGC~4151, NGC~4261, and in particular in NGC~1052 where $\nu_{\rm thick}$ can be estimated at three positions which cover a factor of four in distance (Section~\ref{sec:obs_ff_abs}).  

One should note that there are no free parameters involved in the analysis of the objects above, 
as $\nu_{\rm thick}(r)$ is uniquely set by $\Lbol$ and the absorber distance $r$, both directly 
observed (up to projection effect corrections, likely of order unity). The match of the predicted and observed $\nu_{\rm thick}(r)$ indicates the following: 1.\ the absorbing gas is photoionized; and 
2.\ the absorbing
gas density is set by the incident radiation pressure, i.e. it is RPC gas. 

The results above add to earlier evidence on that RPC sets the gas density in AGN.
Specifically, the NLR density radial dependence and its
ionization structure \citep{D02, paperI, Davies16}, the BLR emission radial dependence \citep{paperII, Netzer20}, 
the Broad Absorption Lines ionization structure \citep{paperIV, MM19}, 
the X-ray lines absorption measure distribution \citep{paperIII}, the X-ray lines
emission measure distribution \citep{paperV}, and galactic scale winds \citep{Stern16, MasRibas19, 
Somalwar20}. These results suggest that generally 
there is no additional significant confining mechanism for ionized gas in AGN, such as ambient 
magnetic or thermal pressure.

One can therefore use the detection of free-free absorption in the radio 
to derive the distance of the absorbing medium from the centre. This method should
apply for gas on scales $\sim 0.1-100$~pc, where one expects $\nu_{\rm thick}\sim 100 - 0.1$~GHz.

\subsection{Emission}

Detection of free-free emission in AGN at the cm regime ($<30$~GHz) is challenging, as this regime
is generally dominated by synchrotron emission. 
It is likely detected in Cygnus~A \citep{C19}, and was suggested in NGC~1068 \citep{G04}.
Both sources are bright in the radio, though NGC~1068 is RQ, but is nearby and luminous. 
High S/N and high resolution
imaging allows to  detect the relatively faint free-free emission, which at the mJy flux level,
close to the non thermal emission, which is a factor of $10^3-10^5$ brighter in the cm regime.
Below we discuss several expected properties of the free-free emission and compare them to observations.

\subsubsection{Predicted strength}
What is the expected contribution of free-free emission to the observed
radio emission in the cm range?  The observed radio loudness parameter, $R=f_\nu(4400\text{\AA})/f_\nu(5~{\rm GH})$,  is $R\sim 0.1-1$
for most RQ AGN   \citep{K89}. Adopting the mean AGN luminosity ratio
$L_{\rm bol}/\nu L_\nu (4400\text{\AA})= 8$ \citep{richards_etal06},  
implies $L_\nu (4400\text{\AA})=1.8\times 10^{-16}L_{\rm bol}$. The RPC free-free solution yields
a maximal contribution of $L_\nu^{\rm ff} (5~{\rm GH})=5\times 10^{-18}L_{\rm bol}$ (Fig.~\ref{fig:ill_side_nu}). Thus,
the radio loudness due to 
free-free emission is only $R^{\rm ff}({\rm 5~GHz})= 0.028$.  The typical
cm radio emission observed in RQ AGN is clearly not produced by free-free emission. However, a fair
fraction ($\sim 20$ per cent) of RQ AGN have $R<0.1$ \citep[fig.~5]{K89}, in particular
the so-called radio silent AGN \citep[e.g.][]{Chiara19}. In such objects, free-free emission can 
provide significant contribution, or possibly even dominate the cm radio emission.

The above $R^{\rm ff}({\rm 5~GHz})$ estimate assumes $\Omega=0.3$ for the NLR free-free emitting gas.
A more accurate estimate of the expected NLR free-free emission can be derived from \LOIII\ (Fig.~\ref{fig:R_vs_OIII}). The \OIII\ line is expected to be a particularly good predictor of the
NLR free-free radio emission since the \OIII\ line is produced at $r/\rdust>200$,
where the gas densities, induced by RPC, are below the \OIII\ 
critical density \citep[fig.~6]{paperI}. This range overlaps well the range of $r/\rdust$ where the 5~GHz free-free emission
is produced (Fig.~\ref{fig:ill_side_nu}). A remaining free parameter, 
which can significantly affect the \LOIII\ versus $L_{5~{\rm GH}}^{\rm ff}$ relation,
is the gas metallicity, which affects the strength distribution among the different emission lines.
The comparison with the \OIII\ observations, shows that indeed
the observed radio emission is higher than predicted by at least a factor
of $\sim 2-3$ or larger (Fig.~\ref{fig:R_vs_OIII}).

The mean observed radio spectral slope  at $5-8.5$~GHz is $\alpha \sim -0.5$ \citep[e.g.][]{laor_etal19}, while free-free emission produces $\alpha \sim -0.1$. 
Thus, the relative contribution of the free-free emission is expected to
increase at higher frequencies. Indeed, the detection of the free-free
emission in Cygnus~A is made at $18-48$~GHz \citep{C19}. The free-free emission may dominate 
at $\nu> 100$~GHz, if it is produced by dustless gas just outside the BLR, for an underlying continuum 
with $\alpha \sim -0.5$, and radio loudness of $R=0.6$ (Fig.~\ref{fig:obs_SED}). 
The free-free emission could be detectable at $\nu<100$~GHz in objects with
a steeper $\alpha$ and a lower $R$ value.

\subsubsection{Synchrotron or free-free mm emission?}

Radio observations of RQ AGN at $\sim 100$~GHz typically show excess 
emission, compared to the low frequency extrapolation \citep{Doi05, Doi11,behar_etal15,behar_etal18}. In addition to free-free, an alternative mechanism which
may produce the observed excess emission, is a compact optically thick synchrotron source.
The observed spectrum will show a transition
from optically thin steep PL synchrotron emission, to a flat PL optically thick  
emission. Such a transition is indeed expected if the compact synchrotron source is produced 
by a magnetized accretion disc corona \citep{FR93, LB08, ID14, RL16}. Indeed, in NGC~1068
the excess emission observed by ALMA above 100~GHz, which is interpreted here 
as free-free emission from the BLR scale, may instead be produced by optically thick
synchrotron emission \citep{Inoue20}.

How can one determine if the excess high frequency emission is synchrotron
or free-free emission? The simplest diagnostic is the value of $\alpha$.
If $\alpha<-0.1$, as found by \cite{ID18} in NGC~985 and IC~4329A at $100-200$~GHz,
then the free-free interpretation is clearly ruled out. 
In contrast, a slope of $\alpha>-0.1$
can be produced by both optically thick synchrotron and by optically thick free-free emission.

The other major difference between free-free and optically thick synchrotron emission  
is the physical scale of the emitting region. Significant
free-free emission at $\sim 100$~GHz is produced at $r/\rdust> 10$ (Fig.~\ref{fig:ill_side_nu}), while the 
corresponding scale for optically thick 
synchrotron emission  is $r< 300 r_{\rm g}$ \citep[fig.~4]{RL16}, 
which roughly corresponds to $r/\rdust < 0.03$. A flat spectrum synchrotron source is therefore 
expected to be more compact
than a flat spectrum free-free source by a factor of $>300$.

\subsubsection{Resolving the free-free emission region}

What is the expected angular scale of the BLR scale free-free emission?
In a nearby AGN, say at 100~Mpc, with moderate luminosity, say $L_{\rm bol}=10^{44}$~erg~s$^{-1}$, 
where $\rdust=0.02$~pc (eq.~\ref{eq:rdust}), the corresponding angular scales are $>0.4$~mas for free-free
emission, and $<1.3$~$\mu$as for optically thick synchrotron emission. 
These angular scales are well below
the maximal resolution of the ALMA array. The angular resolution is $\Theta=\lambda/D$, and for the ALMA array baseline
of $D=16$~km, and for observations at 100~GHz, the angular resolution is only
$\Theta=0.3/1.6\times 10^6=1.9\times 10^{-7}=38$~mas. 

Can the emission be resolved using the Event Horizon Telescope (EHT)? 
The EHT is working at $\lambda=0.13$~cm, with $D\simeq 10,000$~km, and thus reaches
$\Theta_{\rm EHT}=25~\mu$as, and can resolve well the free-free emission scale. 
Indeed, the EHT resolves
the emission down to $\sim 10 r_{\rm g}$ in M~87 \citep{EHT19}. However, the expected enclosed flux within an angular scale of 
$\Theta=\Theta_{\rm mas}$~mas, of a source with $T_{\rm b}=10^9 T_9$~K, at $\nu=10^{11}\nu_{100~{\rm GHz}}$~Hz is 
\begin{equation}
F_\nu=\Theta^2 \frac{2\nu^2}{c^2}kT_{\rm b}=
7.2\times 10^{-23} \Theta_{\rm mas}^2 T_9\nu_{100~{\rm GHz}}^2\ \ \ {\rm erg~s}^{-1}~{\rm Hz}^{-1}~{\rm cm}^{-2}
\end{equation}
\citep[e.g.][]{RL04}, or
\begin{equation}
F_\nu=7.2 \Theta_{\rm mas}^2\nu_{100~{\rm GHz}}^2T_9\ \ \ {\rm Jy}. \label{eq:f_nu_for_res}
\end{equation}
Thus, a synchrotron source, with say $T_{\rm b}\sim 10^{11}$~K, as expected
for equipartition between the B field and the electron energy densities \citep{Readhead94},
will produce a flux of $F_\nu \sim 0.5$~Jy on the $\Theta_{\rm EHT}$ resolution scale. 
This is indeed the flux observed in M87 by the \cite{EHT19}. 
However, in RQ AGN, where $\Theta<1.2$~$\mu$as (see above) 
a value of $T_{\rm b}\sim 10^{11}$~K gives
$F_\nu \sim 10~\mu$~Jy. So, the synchrotron source will be detectable only
if $T_{\rm b}$ is well above $10^{11}$~K. Free-free emission at 100~GHz, from $r/\rdust> 10$,
is characterised by $T_{\rm b}< 1.5\times 10^4$~K (Fig.~\ref{fig:Tb}), which 
corresponds to a tiny flux of $F_\nu \sim 0.1~\mu$Jy (eq.~\ref{eq:f_nu_for_res}), which is far too weak to be detectable by the EHT. 

Thus, only the more extended NLR-scale free-free emission can be detected and resolved,
assuming the specific array used has a low enough $T_{\rm b}$ detection limit, which allows to detect 
optically thin free-free emission.

\subsubsection{Variability}

An indirect but robust upper limit on the size of the emitting region is variability.
In a typical luminous Seyfert ($L_{\rm bol}=10^{44}$~\ergs),
the 100~GHz free-free emitting region is at $10\rdust\sim 0.2$~pc, so can vary only on a year
time scale. A synchrotron source originates from a region $\sim 300$ times smaller,
and can therefore vary on a single day time scale.

Recurrent 100~GHz observations of a few nearby low luminosity AGN by \cite{Doi11} suggest
possible variability on time scales as short as days. The only long term mm monitoring
campaign of a RQ AGN is available for NGC~7469 \citep{Baldi15, Behar20}, which indicates significant
variability on a time scale of a couple of days. The host subtracted optical continuum 
luminosity of NGC~7469 is $2\times 10^{43}$~erg~s$^{-1}$ \citep{Bentz09}, which indicates
$\Lbol\simeq 2\times 10^{44}$~erg~s$^{-1}$. The observed 100~GHz variability on time scale
of a couple of days clearly excludes free-free emission, which can vary only on year time scale
for this $\Lbol$, but is consistent with the expected size of an optically thick synchrotron
source. Thus, although the spectral slope at $\sim 100-300$~GHz is $\alpha\simeq 0$ 
\citep[fig.~2]{Behar20}, which is consistent with free-free emission, the emitting region 
is far too compact. A similar conclusion is reached based on the observed 
$L_{100~{\rm GHz}}/\Lbol=3\times 10^{-16}\ \ {\rm Hz}^{-1}$, which is a factor $\sim 10$
larger than expected for free-free emission from optically thin dustless gas 
(Fig.~\ref{fig:ill_side_nu}), and must therefore be dominated by synchrotron emission.
The observed high $L_{100~{\rm GHz}}/\LOIII$ luminosity ratio also leads to a similar
conclusion.

\subsubsection{Dilution by dust emission}

How high in frequency can we probe the free-free emission? 
The advantage of going to higher frequencies is that the free-free flux density
either rises steeply (optically thick), or remains nearly constant (optically thin).
The free-free flux drop occurs only in the optical regime or above (Fig.~\ref{fig:obs_SED}).
In contrast, the synchrotron emission drops once the emission becomes
optically thin, which is expected to occur below 1000~GHz in RQ AGN \citep{RL16}. 
Thus, the free-free emission
is more likely to become dominant only at the sub mm regime ($>300$~GHz).
However, the observed mean radio and FIR SED extrapolated 
to the sub mm regime, suggests that the steeply rising dust emission dominates above $\sim 500$~GHz
(Fig.~\ref{fig:obs_SED}). The FIR extrapolation of the AGN contribution 
is subject to significant uncertainty \citep[e.g.][]{Bernhard21}, which adds to the
uncertain host contribution. Despite this somewhat uncertain sharply rising ``dust wall''  
at $\nu > 500$~GHz ($\propto \nu^{3.5}$), it may be possible to identify the nuclear 
free-free and synchrotron emission. The cold dust emission
extends on the host galaxy scale, i.e.\ on arcsec scale, which will be resolved out
using sub mm interferometry, in particular with ALMA. For example, \cite{Gar16} detect a
$\sim$10~mJy source on a scale of $\sim$50~mas in the core of NGC~1068 at
$\sim$430~$\mu$m, while the total flux measured at $\sim$400~$\mu$m in NGC~1068, on 
a scale of 75 arcsec, is $\sim$30~Jy \citep{H77}, i.e. a factor of 3000 brighter. 
Thus, the highest frequency where 
nuclear scale free-free emission can be probed, depends on the highest frequency 
currently available on sub mm interferometry (currently 950~GHz on ALMA).

\subsubsection{NGC 1068}

Above we showed (Section \ref{sec:ngc1068}) that the observed 
flat, $\alpha_{\rm 5-8.4~GHz}=-0.17\pm 0.24$,  compact VLBA scale radio
emission cannot be produced by free-free emission, for the following
reasons. First, the observed $T_{\rm b}$ is too high compared to
the expected value for photoionized gas, of either RPC or uniform density gas.
Second, this source lies behind a free-free absorbing screen, and 
the absorption corrected slope is too steep to agree with 
free-free emission. 

We then find that the free-free emission of the absorbing screen, which 
resides at the NLR scale ($r/\rdust\simeq 1000$), is consistent with the observed diffuse emission 
on the same scale, both in amplitude and spectral slope. One should
note, however, that separating out the true diffuse emission from the compact 
synchrotron sources, is subject to some uncertainty. Thus, the detection of the
free-free emission from the NLR-scale free-free absorbing screen is only
tentative.

A potentially more exciting result is the possible detection of free-free emission 
from photoionized gas on the BLR scale, which can produce the 
excess emission observed by ALMA above 100~GHz (Section~\ref{sec:BLR}).
The implied physical scale, based on the RPC results for dustless gas, 
is $r/\rdust=3-4$, with a small covering factor of $\Omega=0.025-0.03$.
Dusty gas, subject to RPC, gives a smaller region  $r/\rdust=1.1-1.7$ with 
a larger $\Omega=0.23-0.28$. 
Since $\rdust=0.04-0.14$~pc (Section~\ref{sec:ngc1068}), the absolute size 
of the free-free emitting region is $r=0.12-0.56$~pc for dustless gas, and
$r=0.044-0.24$~pc for dusty gas. 

The recent $K$-band interferometry of NGC~1068 with GRAVITY \citep{Gravity20},
reveals a flat ring-like structure, with a radius $r=0.24\pm 0.03$~pc, and a
scale height $h/r<0.14$. The RPC results for dustless gas, $r=0.34\pm 0.24$~pc, 
$h/r=0.14\pm 0.1$  is consistent with the GRAVITY results, both in size and covering factor.

The NIR SED in NGC~1068, and generally in AGN, clearly shows that 
the $K$-band emission is produced by hot dust. Why do the RPC results suggest
that the photoionized free-free emitting gas is dustless? A likely explanation is the nature
of the dust. As pointed out in \cite{BL18}, only large graphite grains survive 
the direct illumination of the AGN at $r/\rdust\sim 1$. The UV absorption
cross sections of dusty gas composed of large grains is significantly reduced
\cite[fig.~7]{BL18}, so the depth of the photoionized layer and the associated 
free-free emission, remain similar to that of dustless gas. The large dust grains are
heated by the UV (although they do not dominate the UV opacity) and cool by NIR emission, 
producing the $K$-band emission.

An alternative interpretation of the excess mm emission above 100~GHz in
NGC~1068 is the contribution of an optically thick synchrotron emission \citep{Inoue20}, 
which resides on the innermost accretion disc scales \citep{RL16},
and is likely produced within the accretion disc corona \citep{FR93, LB08}.
The implied size is $\sim$ a light day, compared to $\sim$ a light year
for the free-free emission. In addition, one expects some relation between the 
mm synchrotron emission and the X-ray emission, as both can be produced in the same region.
Thus, variability studies can exclude the free-free origin of the mm
excess in NGC~1068. 

The advantage of the BLR free-free interpretation, is that it naturally produces
a spectral turnover at $\nu$ just below $\sim 1000$~GHz, as observed in NGC~1068; in contrast 
with the synchrotron interpretation, where the turnover depends on the outer radius
of the magnetised plasma cloud, and there is no obvious preference for a
specific radius.

In principle, interferometry at $\nu> 1000$~GHz can determine if the emission slope
remains flat, as expected for optically thin free-free emission, or becomes steep 
since a synchrotron source is expected to become optically thin \citep{RL16}.
However, a THz interferometry is not yet available, so this test cannot be made.

\section{Conclusions} \label{sec:conclusions}

Ionizing radiation likely sets the ionization, temperature and density
structure of the gas it is incident upon, and therefore the line and continuum
emission of the gas. Following earlier studies in this series, on the 
emission and absorption properties of radiation pressure compressed photoionized 
gas, here we focus on the radio regime, in particular the free-free emission and 
absorption of RPC gas in AGN. We find the following:
\begin{enumerate}

\item The distance of a free-free absorbing gas screen from an ionizing
source of a given luminosity, can be found based on the frequency where the
absorber becomes optically thick. The
validity of the distance derived with this method is verified in a few objects where
the free-free absorber is spatially resolved.

\item Free-free emission from gas at the NLR, at the kpc scale, is self-absorbed
below a few hundred MHz. On the hundred pc scale, self-absorption sets in below a few GHz, 
and at the BLR scale free-free self-absorption sets in below a few hundred GHz.

\item The peak of the free-free emission, which occurs slightly above the 
self-absorption frequency, is characterised by $T_{\rm b}<2\times10^4$~K
for free-free emission at all distances.

\item Radio emission with $T_{\rm b}>10^6$~K, as observed
in NGC~1068, cannot be produced by free-free emission from hot gas, 
as the free-free emission of such hot gas over predicts the observed
X-ray emission.

\item Free-free emission produces a radio loudness parameter of
$R^{\rm ff}({\rm 5~GHz})\sim 0.03$, while the typical observed 
value is $R\sim 0.1-1$. Thus, free-free emission may be significant
at 5~GHz only in RQ AGN where $R< 0.1$. 

\item The typical contribution of free-free emission in the radio is expected to rise with
frequency, and may dominate at $\nu>100$~GHz in some RQ AGN.
In particular, if the free-free emission originates in dustless gas just outside the BLR.

\item The excess mm emission observed in NGC~1068 above 100~GHz, may
be produced by photoionized gas at a distance of $r=0.12-0.56$~pc.
This size matches the size of hot dust ring, spatially resolved by GRAVITY, 
at $r=0.24$~pc.

\item Excess mm emission above 100~GHz can also be produced by a compact synchrotron 
source at the innermost accretion disc. It differs dramatically in size 
from a free-free source, being $\sim 300$ times smaller,
and can therefore be differentiated based on the variability time scale.

\item Optimal detection of free-free emission is through mm and sub mm
interferometry, using an array which can detect emission at
$T_{\rm b}< 10^4$~K, such as ALMA. Such observations can map the spatial
distribution of warm photoionized gas in AGN on the sub pc scale, even if this
region is highly absorbed in the optical and in the X-ray regimes.

\end{enumerate}

The improved sensitivity of sub mm interferometers now allows to study RQ AGN. 
Apart from studies of molecular gas emission lines, such observations
can also probe the presence of warm photoionized gas on the sub pc scale, 
and potentially also detect non thermal emission from the innermost accretion disc scale
\citep{Panessa19}. 
	
	\section*{Acknowledgements}
	We thank the referee, G.\ Bicknell, for a thorough and careful review 
	which improved the paper. AB thanks the hospitality of the Physics Department, Technion -- Israel Institute of Technology, where this work had been conceived during his stay as a post-doctoral fellow in 2016. This research was supported by the Israel Science Foundation
(grant no.\ 1008/18).

\section*{Data availability}
The data underlying this article will be shared on reasonable request to the corresponding author.

	\bsp	
	\label{lastpage}
\end{document}